\documentclass[conference]{IEEEtran}
\usepackage{color}
\usepackage{algorithm,algorithmicx,algcompatible, cite}
\usepackage{verbatim, subfigure}
\usepackage{array, calc, multicol}
\usepackage{xspace}
\usepackage{graphics, epstopdf, graphicx, epsfig, psfrag, epsf, subfigure}
\usepackage{amssymb, amsfonts, amsmath, times}
\usepackage{multicol,balance, verbatim}
\usepackage{textcomp}
\newcommand{\ie}{{\em i.e., }}
\newcommand{\Ie}{{\em I.e., }}

\newtheorem{theorem}{Theorem}

\newtheorem{example}{Example}

\DeclareGraphicsExtensions{.eps, .pdf, .png, .jpg}   % optional

\newcommand{\Sset}{\mathcal{S}}
\newcommand{\Kset}{\mathcal{K}}

\newcommand{\Nset}{\mathcal{N}}

\newcommand{\Qset}{\mathcal{Q}}

\DeclareMathOperator*{\argmax}{arg\,max}

\IEEEoverridecommandlockouts

\makeatletter
\newcommand{\oset}[2]{%
{\mathop{#2}\limits^{\vbox to -.5\ex@{\kern-\tw@\ex@
\hbox{\scriptsize #1}\vss}}}}
\makeatother
\begin{document}

%\title{FIFO Queues: Stability Region, Flow Control, and Scheduling}
%\vspace{-80pt}
\title{Flow Control and Scheduling for Shared FIFO Queues over Wireless Networks \vspace{-5pt}  }

%\author{Hulya Seferoglu\\
%%{\small ECE Department}\\
%{\small University of Illinois at Chicago}\\
%{ \small \tt hulya@uic.edu}\\
%\and
%Erdem Koyuncu\\
%%{\small CPCC}\\
%{\small University of California, Irvine}\\
%{ \small \tt ekoyuncu@uci.edu }\\
% \and 
%Shanyu Zhou\\
%%{\small ECE Department}\\
%{\small University of Illinois at Chicago}\\
%{ \small \tt szhou45@uic.edu }\\
%}

\author{
\authorblockN{Shanyu Zhou}
\authorblockA{University of Illinois at Chicago \\
  \tt szhou45@uic.edu}
 \and
 \authorblockN{Hulya Seferoglu}
\authorblockA{University of Illinois at Chicago \\
 \tt hulya@uic.edu}
 \and
\authorblockN{Erdem Koyuncu}
\authorblockA{University of California, Irvine\\
 \tt  ekoyuncu@uci.edu}
}

\maketitle

%\linespread{0.993}

%\vspace{-500pt}
%\vfill

{$\hphantom{a}$}\vspace{-35pt}{}

\allowdisplaybreaks

\begin{abstract}
We investigate the performance of First-In, First-Out (FIFO) queues over wireless networks. We characterize the stability region of a general scenario where an arbitrary number of FIFO queues, which are served by a wireless medium, are shared by an arbitrary number of flows. In general, the stability region of this system is non-convex. Thus, we develop a convex inner-bound on the stability region, which is provably tight in certain cases. The convexity of the inner bound allows us to develop a resource allocation scheme; $dFC$. Based on the structure of $dFC$, we develop a stochastic flow control and scheduling algorithm; $qFC$. We show that $qFC$ achieves optimal operating point in the convex inner bound. Simulation results show that our algorithms significantly improve the throughput of wireless networks with FIFO queues, as compared to the well-known queue-based flow control and max-weight scheduling. 
\end{abstract}

\section{Introduction}\label{sec:intro}

The recent growth in mobile and media-rich applications continuously increases the demand for wireless bandwidth, and puts a strain on wireless networks \cite{cisco_index}, \cite{ericsson_report}. This dramatic increase in demand poses a challenge for current wireless networks, and calls for new network control mechanisms that make better use of scarce wireless resources. Furthermore, most existing, especially low-cost, wireless devices have a relatively rigid architecture with limited processing power and energy storage capacities that are not compatible with the needs of existing theoretical network control algorithms. One important problem, and the focus of this paper, is that low-cost wireless interface cards are built using First-In, First-Out (FIFO) queueing structure, which is not compatible with the per-flow queueing requirements of the optimal network control schemes such as backpressure routing and sheduling \cite{tass_eph1}.

The backpressure routing and scheduling paradigm has emerged from the pioneering work \cite{tass_eph1}, \cite{tass_eph2}, which showed that, in wireless networks where nodes route and schedule packets based on queue backlogs, one can stabilize the queues for any feasible traffic. It has also been shown that backpressure can be combined with flow control to provide utility-optimal operation \cite{neely_mod}. Yet, backpressure routing and scheduling require each node in the network to construct per-flow queues. %To better illustrate this key point, let us first discuss the operation of max-weight scheduling in the following example. Let us consider the following example to further explain the operation of IDNC.
The following example demonstrates the operation of backpressure. % routing and scheduling.

\begin{figure}[t!]
\vspace{-20pt}
\centering
\subfigure[Per-flow queues]{ \label{fig:intro_example_a} \scalebox{.5}{\includegraphics{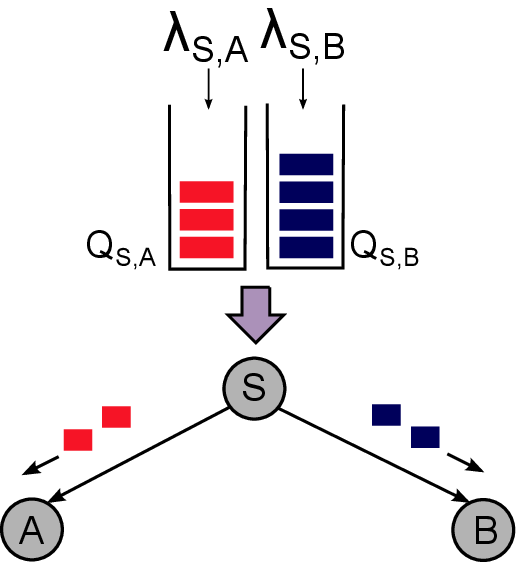}} }
\subfigure[FIFO queue] {\label{fig:intro_example_b} \scalebox{.5}{\includegraphics{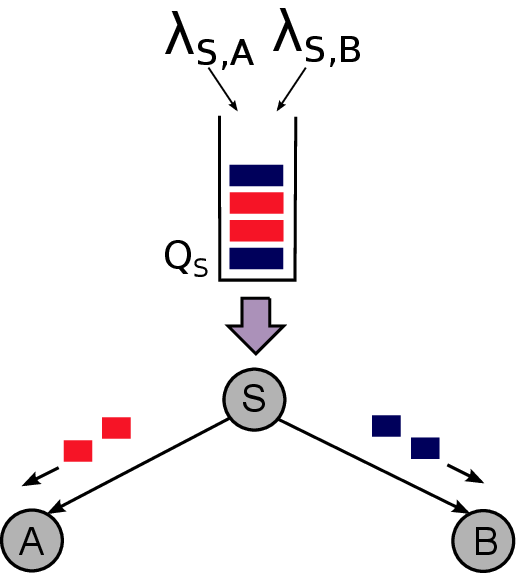}} }
%\subfigure[Per-flow queues]{ \label{fig:intro_example_a} \scalebox{.56}{\includegraphics[bb=0 0 177 198]{figs/FIFO_example_fig_a.eps}} }
%\subfigure[FIFO queue] {\label{fig:intro_example_b} \scalebox{.56}{\includegraphics[bb=0 0 177 198]{figs/FIFO_example_fig_b.eps}} }
\vspace{-5pt}
\caption{Queueing structure of one-hop downlink topology with (a) per-flow queues, and (b) a FIFO queue. %(c) Corresponding stability region assuming that links are in the ON or OFF state with probability $p_1$ and $p_2$ for links $i-j$ and $i-k$, respectively.
} \label{fig:example_fifo_v1}
\vspace{-15pt}
\end{figure}

\begin{example}
Let us consider a canonical example in Fig. \ref{fig:example_fifo_v1}(a), where a transmitter node $S$, and two receiver nodes $A$, $B$ form a one-hop downlink topology. There are two flows with arrival rates $\lambda_{S,A}$ and $\lambda_{S,B}$ destined to nodes $A$ and $B$, respectively. The throughput optimal backpressure scheduling scheme, also known as max-weight scheduling, assumes the availability of
per-flow queues $Q_{S,A}$ and $Q_{S,B}$ as seen in Fig. \ref{fig:example_fifo_v1}(a), and 
%requires the ability to transmit packets based on the flow to which they belong. \Eg the backpressure scheduling algorithm, also known as max-weight scheduling, 
makes a transmission decision at each transmission opportunity based on queue backlogs, \ie $Q_{S,A}$ and $Q_{S,B}$. In particular, the max-weight scheduling algorithm determines $F^{*} = \argmax_{F \in \{A,B\}} Q_{S,F}$, and transmits from queue $Q_{S,F^{*}}$. It was shown in \cite{tass_eph1}, \cite{tass_eph2} that if the arrival rates $\lambda_{S,A}$ and $\lambda_{S,B}$ are inside the stability region of the wireless network, the max-weight scheduling algorithm stabilizes the queues. On the other hand, in some devices, per-flow queues cannot be constructed. In such a scenario, a FIFO queue, say $Q_{S}$ is shared by flows $A$ and $B$ as shown in Fig.~\ref{fig:example_fifo_v1}(b), and the packets are served from $Q_{S}$ in a FIFO manner.
\hfill $\Box$
\end{example}

Constructing per-flow queues may not be feasible in some devices especially at the link layer due to rigid architecture, and one FIFO queue is usually shared by multiple flows. For example, although current WiFi-based devices have more than one hardware queue \cite{madwifi_chipsets}, their numbers are restricted (up to 12 queues according to the list in \cite{madwifi_chipsets}), while the number of flows passing through a wireless device could be significantly higher. Also, multiple queues in the wireless devices are mainly constructed for prioritized traffic such as voice, video, etc., which further limits their usage as per-flow queues. 
On the other hand, constructing per-flow queues may not be preferable in some other devices such as sensors or home appliances for which maintaining and handling per-flow queues could introduce too much processing and energy overhead. Thus, some devices, either due to rigid architecture or limited processing power and energy capacities, inevitably use shared FIFO queues, which makes the understanding of the behavior of FIFO queues over wireless networks very crucial.
%if we are required to use shared FIFO queues either due to rigid architecture or limited processing power and energy capacity, 
%In these examples, although there could be trade-off

{\em Example 1 - continued:}
Let us consider Fig.~\ref{fig:example_fifo_v1} again. When a FIFO queue is used instead of per-flow queues, the well-known head-of-line (HOL) blocking phenomenon occurs. As an example, suppose that at transmission instant $t$, the links $S-A$ and $S-B$ are at ``ON'' and ``OFF'' states, respectively. In this case, a packet from $Q_{S,A}$ can be transmitted if per-flow queues are constructed. Yet, in FIFO case, if HOL packet in $Q_S$ belongs to flow $B$, no packet can be transmitted and wireless resources are wasted.
\hfill $\Box$

Although HOL blocking in FIFO queues is a well-known problem, achievable throughput with FIFO queues in a wireless network is generally not known. In particular, stability region of a wireless network with FIFO queues as well as resource allocation schemes to achieve optimal operating points in the stability region are still open problems.
%the impact of FIFO queues to  wireless networks and corresponding scheduling policies are generally not known. In particular,

In this work, we investigate FIFO queues over wireless networks. We consider a wireless network model presented in Fig.~\ref{fig:main-example} with multiple FIFO queues that are in the same transmission and interference range. (Note that this scenario is getting increasing interest in practice in the context of device-to-device and cooperative networks \cite{microcast}.) 
Our first step towards understanding the performance of FIFO queues in such a setup is to characterize the stability region of the network. Then, based on the structure of the stability region, we develop efficient resource allocation algorithms; {\em Deterministic FIFO-Control} ($dFC$) and {\em Queue-Based FIFO-Control} ($qFC$). The following are the key contributions of this work:
\begin{itemize}
\item We characterize the stability region of a general scenario where an arbitrary number of FIFO queues are shared by an arbitrary number of flows.
\item The stability region of the FIFO queueing system under investigation is non-convex. Thus, we develop a convex inner-bound on the stability region, which is provably tight for certain operating points.
\item We develop a resource allocation scheme; $dFC$, and a queue-based stochastic flow control and scheduling algorithm; $qFC$. We show that $qFC$ achieves optimal operating point in the convex inner bound.
\item We evaluate our schemes via simulations for multiple FIFO queues and flows. The simulation results show that our algorithms significantly improve the throughput as compared to the well-known queue-based flow control and max-weight scheduling schemes.
\end{itemize}

The structure of the rest of the paper is as follows. Section~\ref{sec:system} gives an overview of the system model. Section~\ref{sec:stability_region} characterizes the stability region with FIFO queues. Section~\ref{sec:oFC_qFC} presents our resource allocation algorithms; $dFC$ and $qFC$. Section~\ref{sec:performance} presents simulation results. Section~\ref{sec:related} presents related work. Section~\ref{sec:conclusion} concludes the paper.

\section{System Model}\label{sec:system}
\begin{figure}
\vspace{5pt}
\centering
\scalebox{.55}{\includegraphics{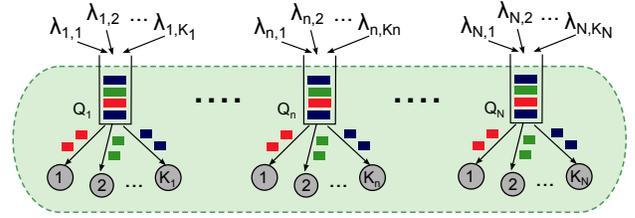}}
\vspace{-5pt}
\caption{The wireless network model that we consider in this paper. $N$ FIFO queues share a wireless medium, where the $n$th FIFO queue,  $\Qset_{n}$ carries $\Kset_{n}$ flows towards their respective receiver nodes. The arrival rate of the $k$th flow passing through the $n$th queue is $\lambda_{n,k}$.}
\label{fig:main-example}
\vspace{-15pt}
\end{figure}

{\em Wireless Network Setup:}
We consider a wireless network model presented in Fig.~\ref{fig:main-example} with $N$ FIFO queues. Let $\Nset$ be the set of FIFO queues, $\Qset_n$ be the $n$th FIFO queue, and $\Kset_n$ be the set of flows passing through $\Qset_n$. Also, let $Q_{n}$ and $K_n$ denote the cardinalities of sets $\Qset_n$ and $\Kset_n$, respectively. We assume in our analysis that time is slotted, and $t$ refers to the beginning of slot $t$.

{\em Flow Rates:}  Each flow passing through $\Qset_n$ and destined for node $k$ is generated according to an arrival process $\lambda_{n,k}(t)$ at time slot $t$. The arrivals are i.i.d. over the time slots such that for every $n \in \Nset$ and $k \in \Kset_{n}$, we have $\lambda_{n,k} = E[\lambda_{n,k}(t)]$ and $E[\lambda_{n,k}(t)^{2}]<\infty$, where $E[\cdot]$ denotes the expected value.

{\em Channel Model:} In our setup in Fig.~\ref{fig:main-example}, as we mentioned earlier, we assume that all FIFO queues are in the same transmission and interference range, \ie only one FIFO queue could be served by a shared wireless medium at time $t$. On the other hand, a channel state from a FIFO queue to a receiver node may vary. In particular, at slot $t$, $\boldsymbol C(t) = \{C_{n,k}(t)\}_{\forall n \in \Nset, k \in \Kset_{n}}$ is the channel state vector, where $C_{n,k}(t)$ is the state of the link at time $t$ from the $n$th queue $\Qset_{n}$ to receiver node $k$ such that $k \in \Kset_{n}$.  The link state $C_{n,k}(t)$ takes values from the set $\{ON,OFF\}$ according to a probability distribution which is i.i.d. over time slots. If $C_{n,k}(t) = ON$, packets can be transmitted to receiver node $k$ with rate $R_{n,k}$. We assume, for the sake of simplicity in this paper, that $R_{n,k}=1$, and $1$ packet can be transmitted at time slot $t$ if $C_{n,k}(t) = ON$. %\footnote{It is straightforward to extend our results for larger values of $R_{n,k}$. Basically, time slots are increased or decreased based on $R_{n,k}$ so that only one packet could be transmitted at a slot. On the other hand, different values of $R_{n,k}$, \ie the case $R_{n,k} \neq R_{n',k'}$, are not straightforward, and left for the future work.} 
If $C_{n,k}(t) = OFF$, no packets are transmitted. The $ON$ and $OFF$ probabilities of $C_{n,k}(t)$ are $\bar{p}_{n,k}$ and $p_{n,k}$, respectively. Note that $C_{n,k}(t)$ only determines the channel state; \ie the actual transmission opportunity from $\Qset_n$ depends on the HOL packet as explained next.

{\em Queue Structure and Evolution:}  Suppose that the Head-of-Line (HOL) packet of $\Qset_{n}$ at time $t$ is $H_{n}(t)\in \Kset_{n}$. The HOL packet together with the channel state defines the state of $\Qset_{n}$. In particular, let $S_n(t)$ be the state of $\Qset_{n}$ at time $t$ such that $S_n(t) \in \{ON,OFF\}$. The state of $\Qset_{n}$ is $ON$, \ie $S_{n}(t) = ON$ if $C_{n,H_{n}(t)} = ON$ at time $t$. Otherwise, $S_{n}(t) = OFF$. We define $\Sset$ $=$ $\{ (S_1,$ $\ldots,$ $S_N)$ $|$ $S_1,$ $\ldots,$ $S_N$ $\in$ $\{ON,OFF\} \}$ as the set of the states of all FIFO queues.

Let us now consider the evolution of the HOL packet. If the state of queue $\Qset_{n}$ is $ON$ at time $t$, \ie $S_{n}(t)=ON$, the HOL packet can be transmitted (depending on the scheduling policy). If we assume that HOL packet is transmitted according to the scheduling policy, then a new packet is placed in the HOL position in $\Qset_{n}$. The probability that this new HOL packet belongs to the $k$th flow is $\alpha_{n,k}$ and it depends on the arrival rates via $\alpha_{n,k} = \frac{\lambda_{n,k}}{\sum_{k \in \Kset_{n}} \lambda_{n,k}}$.

Now, we can consider the evolution of $\Qset_{n}$. At time $t$, $\sum_{k \in \Kset_{n}} \lambda_{n,k}(t)$ packets arrive to $\Qset_{n}$, and $g_{n}(t)$ packets are served according to the FIFO manner. Thus, queue size $Q_{n}(t)$ evolves according to the following dynamics.
\begin{align} \label{eq:queue_Qn}
& Q_{n}(t+1) \leq \max [Q_{n}(t) - g_{n}(t), 0] + \sum_{k \in \Kset_{n}} \lambda_{n,k}(t).
\end{align} 
Note that $g_{n}(t)$ depends on the states of the queues; $\Sset(t)$ at time $t$, which characterize the stability region of the wireless network.  Note that $\Sset(t)$ depends on arrival rates of flows to each FIFO queue; \ie $\lambda_{n,k}$ as well as the $ON$-$OFF$ probability of each link, \ie $p_{n,k}$. In the next section, by taking into account $\lambda_{n,k}$ and $p_{n,k}$, we characterize the stability region of the wireless network.

\section{Stability Region} \label{sec:stability_region}
\begin{figure}
%\vspace{-15pt}
\centering
%\scalebox{.58}{\includegraphics[bb=0 0 177 198]{figs/FIFO_one_queue_fig.eps}}
\subfigure[FIFO queue]{ \label{fig:FIFO_one_queue_fig_a} \scalebox{.6}{\includegraphics{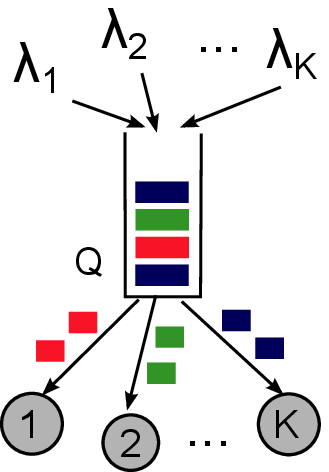}} } \hspace{5pt}
\subfigure[Stability region]{ \label{fig:FIFO_one_queue_fig_b} \scalebox{.6}{\includegraphics{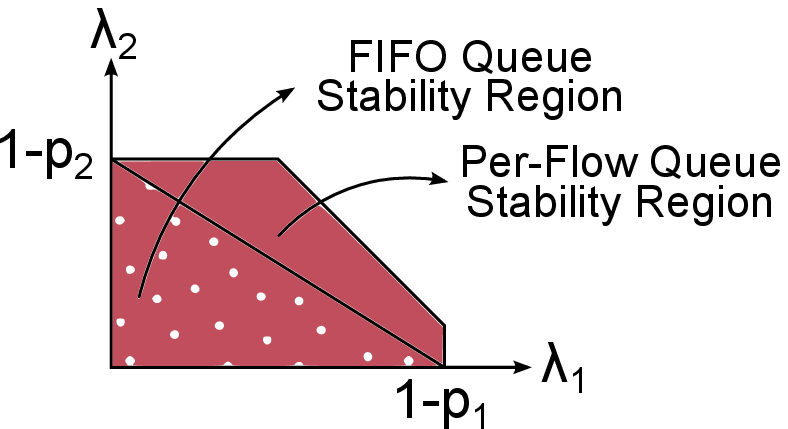}} }
\vspace{-10pt}
\caption{(a) Single-FIFO queue; $\Qset$ is shared by $K$ flows. (b) Stability region of a single-FIFO queue as well as per-flow queues with two flows.}
\label{fig:FIFO_one_queue_fig}
\vspace{-5pt}
\end{figure}

In this section, our goal is to characterize the stability region of a wireless network where an arbitrary number of FIFO queues are served by a wireless medium. We first begin with the single-queue case shown in Fig.~\ref{fig:FIFO_one_queue_fig} to convey our approach for a canonical scenario, then we extend our stability region analysis for arbitrary number of FIFO queues and flows.

\subsection{Single-FIFO Queue} \label{sec:stability_single_queue}
We study the special case of a single FIFO queue $\mathcal{Q}_1$ where $\mathcal{N} = \{1\}$ with $N=1$. For this special case, we thus drop the queue index $n$ from the notation in Section \ref{sec:system} for brevity. In other words, we write $\mathcal{Q}$ instead of $\mathcal{Q}_n$, $C_{k}(t)$ instead of $C_{n,k}(t)$, and so on. Our main result in this context is then the following theorem.
\begin{theorem}\label{theorem1}
For a FIFO queue $\Qset$ shared by $\Kset$ $=$ $\{1,$ $\ldots$ $,K\}$ flows, if the channel states $C_{k}(t)$ and arrival rates $\lambda_{k}(t)$ are i.i.d. over time slots, the stability region $\Lambda$ includes all arrival rates satisfying
\begin{align} \label{eq:stab_one_queue}
\sum_{k \in \Kset} \frac{\lambda_{k}}{\bar{p}_{k}} \leq 1.
\end{align} In other words, the stability region of the single-FIFO queue system is $\Lambda = \{ \{\lambda_{k}\}_{k \in \Kset} |$ (\ref{eq:stab_one_queue})$, \lambda_{k} \geq 0, \forall k \in \Kset\}$.
\end{theorem}
{\em Proof:} The state of the FIFO queue $\Qset$ takes values from $\{ON,OFF\}$ depending the HOL packet and the states of the wireless links. Now, let us take a closer look at the FIFO states. The $OFF$ state occurs if for some $k\in\mathcal{K}=\{1,\ldots,K\}$ we have $H = k$ and $C_{k} = OFF$. Let $z_k$ be the state that $H = k$ and $C_{k} = OFF$. We denote the probability of $z_k$ as $P[z_k] = P[H=k,\,C_k = OFF]$. Also, let $z_0$ be the state that FIFO queue is at $ON$ state for some HOL packet. The state $z_0$ happens precisely when the channel corresponding to the HOL packet is in the $ON$ state. Therefore, the probability of $z_0$ is $P[z_0] = P [C_H = ON]$.
%where $H_n = ANY$ means that the HOL packet of the queue may belong to any flow in $\Kset_n$.

Having defined the queue state probabilities, we can observe that the packets from the FIFO queue could be served only at state $z_0$. It is also clear that the sum of the arrival rates to the queue $\Qset$ should be less than the service rate, which is $P[z_0]$. Noting that we assumed $R_{k}=1$, we conclude that $\sum_{k \in \Kset} \lambda_{k} \leq P[z_0]$.

Let us now calculate $P[z_0]$ and $P[z_k],\,k \in \Kset$ using a Markov chain with states; $z_0$ and $z_k,\,k \in \Kset$. We first show that the state transition probability from $z_0$ to $z_k$ is $P_{0,k} \triangleq \alpha_{k} p_{k}$, where $\alpha_k = \frac{\lambda_{k}}{\sum_{k \in \Kset} \lambda_{k}}$. Since we consider only one FIFO queue, when the queue is at state $z_0$, the HOL packet is always transmitted. The new HOL packet in the next state will belong to the $k$th flow with probability $\alpha_{k}$, and $C_{k} = OFF$ with probability $p_{k}$. Therefore, the state transition probability from $z_0$ to $z_k$ is $P_{0,k} = \alpha_{k} p_{k}$, as claimed.

The probability of moving from state $z_k$ to $z_0$ is $P_{k,0} \triangleq \bar{p}_{k}$ as we can move to the unblocking state $z_0$ from the blocking state $z_k$ if the channel is $ON$ (with probability $\bar{p}_{k}$.). On the other hand, staying in the blocking state $z_k$ is the $OFF$ probability of the channel $C_{k}$. Thus, $P_{k,k} \triangleq p_{k}$. Note that the expressions for $P_{k,0}$ and $P_{k,k}$ do not involve the quantity $\alpha_{k}$. The reason is that $z_k$ is the blocking state, so when we move from $z_k$ to another state (or staying at state $z_k$), the HOL packet is not transmitted and does not change (because $C_{k}=OFF$ at state $z_k$).

\begin{figure}
\vspace{5pt}
\centering
\scalebox{.45}{\includegraphics{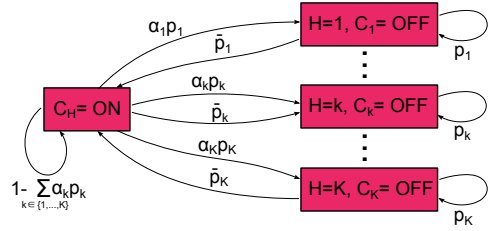}}
%\vspace{-5pt}
\caption{Markov chain for the single-FIFO queue system shown in Fig~\ref{fig:FIFO_one_queue_fig_a}.}
\label{fig:markov_chain_single_fifo}
\vspace{-5pt}
\end{figure}

For any given $k,\,l \in \Kset$ with $k \neq l$, the state transition probability from $z_{k}$ to $z_{l}$ is $P_{k,l} \triangleq 0$. This follows since it is not possible to move from a blocking state to another (the HOL packet cannot be transmitted.). Finally, the probability of staying at state $z_0$ is $P_{0,0} \triangleq 1 - \sum_{k \in \Kset} \alpha_{k}p_{k}$ as the condition $\sum_{k=0}^{K} P_{0,k}=1$ should be satisfied. The state transition probabilities are as shown in Fig.~\ref{fig:markov_chain_single_fifo}.

Now that we know the state transition probabilities of our Markov chain, we can calculate the balance equations, and these yield $P[z_0] = \frac{\sum_{k \in \Kset} \lambda_{k} }{\sum_{k \in \Kset} \lambda_{k}/\bar{p}_{k}}$. The calculations are provided in the following. 

Let $P[\boldsymbol z] = \begin{bmatrix} P[z_0] & P[z_1] & \ldots & P[z_K] \end{bmatrix}^{T}$. In the steady the state, the following set of equations are satisfied for the Markov Chain shown in Fig.~\ref{fig:markov_chain_single_fifo}.  
\begin{align} \label{eq:ststmatrix}
P[\boldsymbol z]^{T}   
\begin{bmatrix}
1 - \sum_{k \in \Kset} \alpha_kp_k & \alpha_1p_1  & \ldots & \alpha_Kp_K \\
\bar{p}_1 						  & p_1          & \ldots & 0 			\\
\bar{p}_2 						  & 0            & \ldots & 0 			\\
   \vdots 						  & \vdots       & \ddots & \vdots 	\\
\bar{p}_{k} 						  & 0            & \ldots & p_K           
\end{bmatrix} 
= P[\boldsymbol z]
\end{align} 
If we combine the $(k+1)$th equation in (\ref{eq:ststmatrix}), which is $P[z_k] = P[z_0] \frac{\alpha_kp_k}{\bar{p}_{k}}$, and the fact that $P[z_0] + \sum_{k \in \Kset} P[z_k] = 1$, we have
\begin{align}
P[z_0] = \frac{\sum_{k \in \Kset} \lambda_{k}}{\sum_{k \in \Kset} \lambda_{k}/\bar{p}_{k}}
\end{align}

We can then obtain $\sum_{k \in \Kset} \lambda_{k} \leq P[z_0] = \frac{\sum_{k \in \Kset} \lambda_{k} }{\sum_{k \in \Kset} \lambda_{k}/\bar{p}_{k}}$ which is equivalent to (\ref{eq:stab_one_queue}). This concludes the proof.
\hfill $\blacksquare$

\begin{example}
Now suppose that single-FIFO queue $\Qset$ is shared by two flows with rates $\lambda_{1}$ and $\lambda_{2}$. According to Theorem~\ref{theorem1}, the arrival rates should satisfy ${\lambda_{1}}/{\bar{p}_{1}} + {\lambda_{2}}/{\bar{p}_{2}} \leq 1$ for stability. This stability region is shown in Fig.~\ref{fig:FIFO_one_queue_fig}(b). In the same figure, we also show the stability region of per-flow queues, \cite{neely_book}. As seen, the FIFO stability region is smaller as compared to per-flow capacity region. Yet, we still need flow control and scheduling algorithms to achieve the optimal operating point in this stability region. This issue will be discussed later in Section~\ref{sec:oFC_qFC}.
\hfill $\Box$
\end{example}

\subsection{Arbitrary Number of Queues and Flows}
We now consider a wireless network with arbitrary number of FIFO queues and flows as shown in Fig.~\ref{fig:main-example}. 
%The main challenge in this setup as compared to the single-FIFO queue case is that packet scheduling decisions affect the stability region. 
The main challenge in this setup is that packet scheduling decisions affect the stability region. 
For example, if both $\Qset_1$ and $\Qset_n$ in Fig.~\ref{fig:main-example} are at $ON$ state, a decision about which queue to be served should be made. This decision affects future transmission opportunities from the queues, hence the stability region.

In this paper, we consider a scheduling policy where the packet transmission probability of each queue depends only on the queue states. In other words, if the state of the FIFO queues is $(S_1, \ldots, S_N) \in \Sset$, a packet from queue $n$ is transmitted with probability $\tau_n(S_1, \ldots, S_N)$. We call this scheduling policy the {\em queue-state} policy.
%\footnote{Note that in general, in addition to queue states, the HOL packet of each queue may also be taken into account for choosing the transmission probabilities, resulting in a {\em queue-state-HOL} policy. In such a scenario, the packet from the $n$th queue would be transmitted with probability, say $\tau_{n,H_n}(S_1, \ldots, S_N, H_1, \ldots, H_N)$. Due to space limitations, the stability analysis of queue-state-HOL policies are provided in our technical report \cite{thisTechRep} for some certain specific topologies. The case of an arbitrary topology appears to be more involved and is left as future work.} 
Note that as $\tau_{n}$ $(S_1,$ $\ldots,$ $S_N)$ is the transmission probability from queue $\Qset_{n}$, we have the obvious constraint
 \begin{align}
 \label{eq:sum_tau}
\sum_{n \in \Nset} \tau_{n}(S_1,\ldots,S_N) \leq 1, \forall (S_1,\ldots,S_{N}) \in \Sset.
\end{align}
Our main result is then the following theorem.
\begin{theorem} \label{theorem2}
For a wireless network with $N$ FIFO queues, if a queue-state policy $\tau_{n}$ is employed, then the stability region consists of the flow rates that satisfy
\begin{align} \label{eq:lamdba_nk}
& \lambda_{n,k} \leq  \sum_{(S_1, \ldots, S_N) \in \Sset} \biggl \{ \frac{\lambda_{n,k} 1_{[S_{n}]}}{\sum_{k \in \Kset_{n}} \lambda_{n,k}/\bar{p}_{n,k}} \nonumber \\
& \prod_{m \in \Nset-\{n\}}  \biggr( \frac{\sum_{k \in \Kset_{m}} \lambda_{m,k} \rho_{m,k}(S_m)}{\sum_{k \in \Kset_{m}} \lambda_{m,k}/\bar{p}_{m,k}} \biggl) \tau_{n} (S_1, \ldots, S_N) \biggr \}, \nonumber \\
& \forall n \in \Nset, k \in \Kset_{n},
\end{align} 
where 
\begin{align*}
1_{[S_{n}]} & = \left\{\begin{array}{rl}1,& S_n = ON \\ 0 , & S_{n} = OFF\end{array}\right., \\
\rho_{m,k}(S_m) & = \left\{\begin{array}{rl} 1, & S_m = ON \\ {p_{m,k}}/{\bar{p}_{m,k}}, & S_m = OFF \end{array}  \right..
\end{align*}
%$1_{[S_{n}]} = 1$ if $S_n = ON$, and $1_{[S_{n}]} = 0$, otherwise. Furthermore, $\rho_{m,k}(S_m) = 1$ if $S_m = ON$, and $\rho_{m,k}(S_m) = {p_{m,k}}/{\bar{p}_{m,k}}$ if $S_m = OFF$.  
%
%As $\tau_{n}(S_1, ..., S_N)$ is the transmission probability from queue $\Qset_{n}$, their sum over all queues should be less than $1$;
% \begin{align}
% \label{eq:sum_tau}
%\sum_{n \in \Nset} \tau_{n}(S_1, ..., S_N) \leq 1, \forall (S_1, ..., S_{N}) \in \Sset
%\end{align}
\end{theorem}
{\em Proof:} The proof is provided in Appendix A. \hfill $\blacksquare$

%The proof of this theorem does not follow from the proof of Theorem \ref{theorem1}, because for arbitrary number of FIFO queues and flows, constructing a Markov chain and providing a solution to the global balance equations for that Markov chain becomes intractable. Thus, we use a different approach, which includes the fact that the distributions of head-of-line (HOL) packets are independent. The details of the proof are provided in Appendix A. %, and the detailed proof is provided in \cite{thisTechRep}.
%\hfill $\blacksquare$

The stability region of a FIFO queue system with $N$ FIFO queues served by a wireless medium is characterized by $\Lambda$ $=$ $\{ \{\lambda_{n,k}\}_{\forall n \in \Nset, k \in \Kset_{n}} |$ (\ref{eq:lamdba_nk}), (\ref{eq:sum_tau})$,$ $\lambda_{n,k}$ $\geq 0,$ $n$ $\in$ $\Nset,$ $k$ $\in$ $\Kset_{n},$  $\tau_{n}(S_1,$ $\ldots,$ $S_N)$ $\geq$ $0,$ $\forall$ $n$ $\in \Nset, (S_1, \ldots, S_{N}) \in \Sset\}$.
%
%Note that in Eq.~(\ref{eq:lamdba_nk}), the term $\frac{\lambda_{n,k} }{\sum_{k \in \Kset_{n}} \lambda_{n,k}/\bar{p}_{n,k}}$ is the probability that HOL packet from $\Qset_{n}$ is $k \in \Kset_{n}$ and the channel $C_{n,k}$ is at the $ON$ state. $1_{[S_{n}]}$ is an indicator whether $S_n = ON$.  On the other hand, $\frac{\sum_{k \in \Kset_{m}} \lambda_{n,k} \rho_{m,k}(S_m)}{\sum_{k \in \Kset_{m}} \lambda_{m,k}/\bar{p}_{m,k}}$ is the probability that the $m^{th}$ queue $\Qset_{m}$ is at state $S_m$. 

\begin{figure}
\vspace{-10pt}
\centering
\subfigure[Two FIFO Queues]{ \label{fig:FIFO_two_queue_fig_a} \scalebox{.40}{\includegraphics{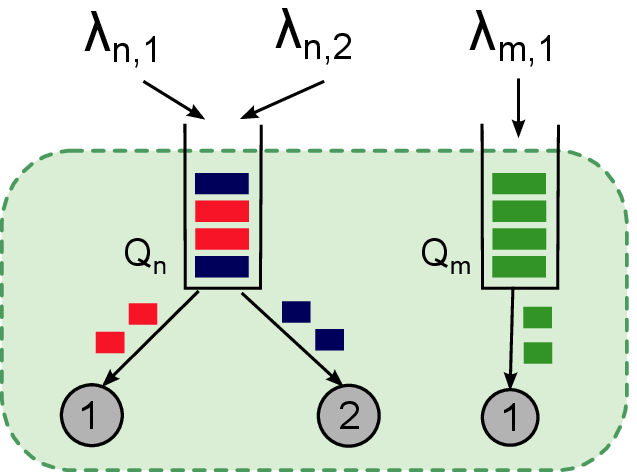}} }
\subfigure[Stability Region]{ \label{fig:FIFO_two_queue_fig_b} \scalebox{.28}{\includegraphics{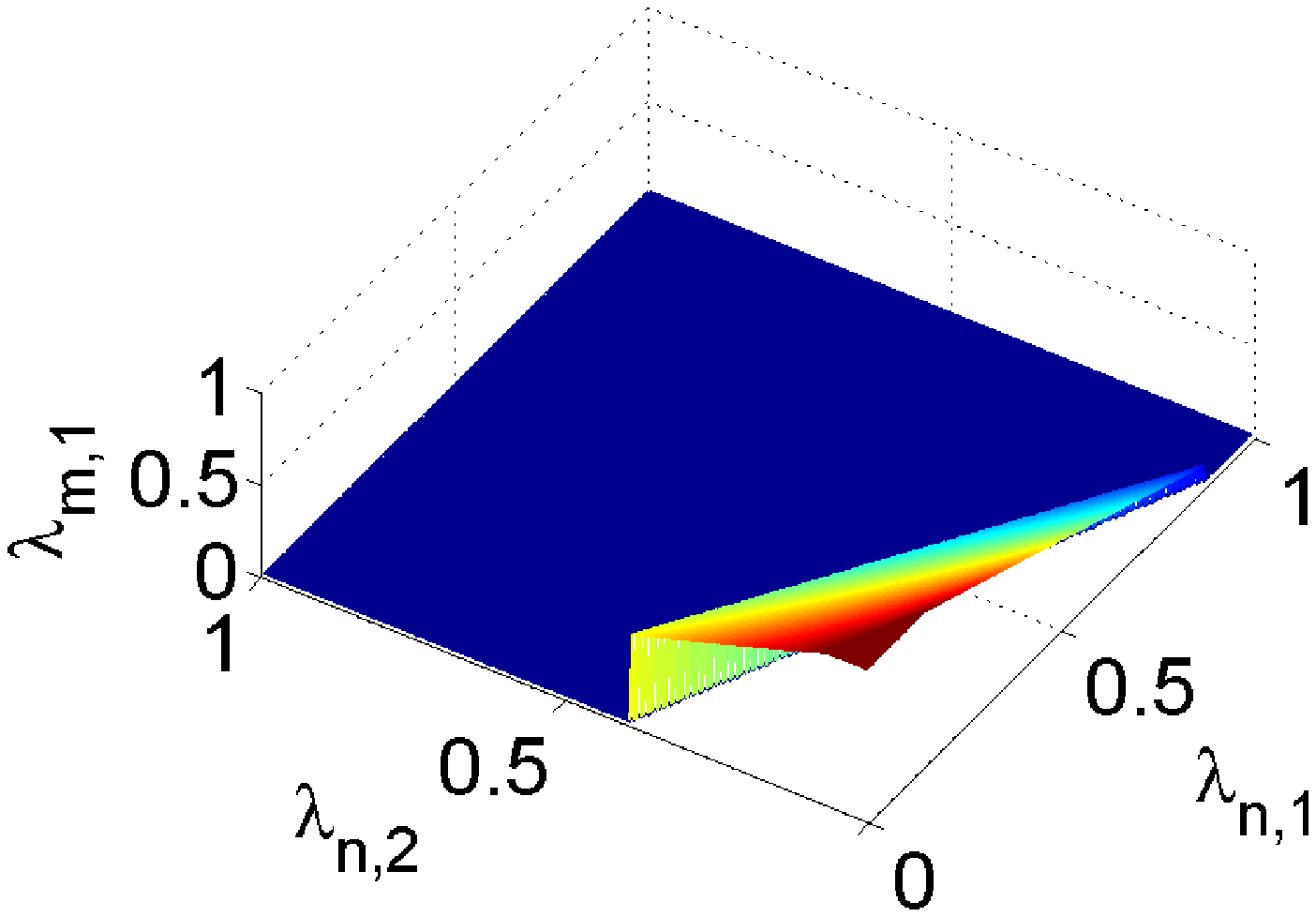}} }
\vspace{-5pt}
\caption{(a) Two FIFO queues; $\Qset_{n}$ and $\Qset_{m}$ are shared by two and one flows, respectively. (b) Three dimensional stability region with $\lambda_{n,1}$, $\lambda_{n,2}$ and $\lambda_{m,1}$ for the two-FIFO queues scenario shown in (a) when $p_{n,1}=0.6$, $p_{n,2}=0.1$, and $p_{m,1}=0.7$.}
\label{fig:two_queue}
\vspace{-15pt}
\end{figure}

\begin{example}
%{\em Example 3:}
Now let us consider two FIFO queues $\Qset_{n}$ and $\Qset_{m}$ which are shared by three flows with rates; $\lambda_{n,1}$, $\lambda_{n,2}$, and $\lambda_{m,1}$ (Fig.~\ref{fig:two_queue}(a)). According to Theorem~\ref{theorem2}, the stability region $\Lambda$ should include arrival rates satisfying inequalities in (\ref{eq:lamdba_nk}) and (\ref{eq:sum_tau}). In this example, with two queues and three flows, these inequalities are equivalent to
\begin{align} \label{eq:two_queue_three_flows}
\lambda_{n,1} + \lambda_{n,2} + \lambda_{m,1} \leq  \bigr( \frac{p_{m,1}(\lambda_{n,1}+\lambda_{n,2})}{\lambda_{n,1}/\bar{p}_{n,1} + \lambda_{n,2}/\bar{p}_{n,2}} \bigl) + \bar{p}_{m,1}
\end{align} with $\lambda_{n,1}/\bar{p}_{n,1}$ $+$ $\lambda_{n,2}/\bar{p}_{n,2}$ $\leq 1$, and $\lambda_{m,1}/\bar{p}_{m,1}$ $\leq$ $1$. The stability region corresponding to these inequalities is the region below the surface  in Fig.~\ref{fig:two_queue}(b). \footnote{Note that the time sharing argument to convexify the stability region does not apply to this scenario, because the non-convexity comes from the relationship among the arrival rates instead of the service rates from the FIFO queues. Thus, the centralized time-sharing for the arrival rates is not practical.}
\hfill $\Box$
\end{example}

In general, we wish to find the optimal operating points on the boundary of the stability region $\Lambda$. However, the stability region may not be convex for arbitrary number of queues and flows. Developing a convex inner bound on the stability region is crucial for developing efficient resource allocation algorithms for wireless networks with FIFO queues. We thus next propose a convex inner bound on the stability region.

\subsection{A Convex Inner Bound on the Stability Region:} \label{sec:stability_single_innerBound}
%In this section, we consider an inner bound on the stability region $\Lambda$. As we mentioned earlier, the stability region $\Lambda$ may not be  convex for arbitrary number of queues and flows, so developing a convex inner bound on the stability region is crucial for developing efficient resource allocation algorithms for wireless networks with FIFO queues.
Let us consider a flow with arrival rate $\lambda_{n,k}$ to the FIFO queue $\Qset_{n}$. If there are no other flows and queues in the network, then the arrival rate should satisfy $\lambda_{n,k}/\bar{p}_{n,k} \leq 1$ according to Theorem~\ref{theorem2}. In this formulation, $\lambda_{n,k}/\bar{p}_{n,k}$ is the total amount of wireless resources that should be allocated to transmit  the flow with rate $\lambda_{n,k}$. For multiple-flow, single-FIFO case, the stability region is $\sum_{k \in \Kset_n} {\lambda_{n,k}}/{\bar{p}_{n,k}} \leq 1$. Similar to the single-flow case, ${\lambda_{n,k}}/{\bar{p}_{n,k}}$ term is the amount of wireless resources that should be allocated to the $k$th flow. Finally, for the general stability region for arbitrary number of queues and flows, let us consider (\ref{eq:lamdba_nk}) again. Assuming $\psi_{m}(S_m) = \frac{\sum_{k \in \Kset_{m}} \lambda_{m,k} \rho_{m,k}(S_m)}{\sum_{k \in \Kset_{m}} \lambda_{m,k}/\bar{p}_{m,k}}$, we can write $\sum_{k \in \Kset_{n}} \lambda_{n,k}$ from (\ref{eq:lamdba_nk}) as;
\begin{align} \label{eq:gec_v1}
& \sum_{k \in \Kset_{n}} \lambda_{n,k} \leq \sum_{(S_1, \ldots, S_N) \in \Sset} \biggl \{ \frac{\sum_{k \in \Kset_{n}} \lambda_{n,k} 1_{[S_{n}]}}{\sum_{k \in \Kset_{n}} \lambda_{n,k}/\bar{p}_{n,k}} \nonumber \\
& \prod_{m \in \Nset-\{n\}}  \psi_{m}(S_m)  \tau_{n} (S_1, \ldots, S_N) \biggr \}, \forall n \in \Nset, k \in \Kset_{n}
\end{align} which, assuming that $\sum_{k \in \Kset_{n}}\lambda_{n,k} > 0$,  is equivalent to
\begin{align} \label{eq:gec_v2}
& \sum_{k \in \Kset_{n}} \lambda_{n,k}/\bar{p}_{n,k} \leq \sum_{(S_1, \ldots, S_N) \in \Sset} 1_{[S_{n}]} \prod_{m \in \Nset-\{n\}}  \psi_{m}(S_m) \nonumber \\
& \tau_{n} (S_1, \ldots, S_N), \forall n \in \Nset, k \in \Kset_{n}
\end{align} Intuitively speaking, the right hand side of (\ref{eq:gec_v2}) corresponds to the amount of wireless resources that is allocated to the $n$th queue $\Qset_{n}$. Thus, similar to the single-FIFO queue, we can consider that ${\lambda_{n,k}}/{\bar{p}_{n,k}}$ term corresponds to the amount of wireless resources that should be allocated to the $k$th flow.

Our key point while developing an inner bound on the stability region is to provide rate fairness across competing flows in each FIFO queue. Since each flow requires ${\lambda_{n,k}}/{\bar{p}_{n,k}}$ amount of wireless resources; it is intuitive to have the following equality ${\lambda_{n,k}}/{\bar{p}_{n,k}} = {\lambda_{n,l}}/{\bar{p}_{n,l}}$, $k \neq l$ to fairly allocate wireless resources across flows. More generally, we define a function $a_n = \lambda_{n,k}/(\bar{p}_{n,k})^{\beta}$, $\forall k \in \Kset_{n}$ where $\beta \geq 1$, and we develop a stability region for $a_n$ instead of $\lambda_{n,k}$. The role of the exponent $\beta$ is to provide  flexibility to the targeted fairness. For example, if we want to allocate more resources to flows with better channels, then $\beta$ should be larger. 

Now, by the definition of  $a_n$, we have the equivalent form
\begin{align} \label{eq:an}
& a_n \leq \sum_{(S_1, \ldots, S_N) \in \Sset} \frac{1_{[S_n]}}{\sum_{k \in \Kset_{n}} (\bar{p}_{n,k})^{\beta-1}} \prod_{m \in \Nset - \{n\}}  \omega_{m}(S_{m}) \nonumber \\
& \tau_{n}(S_1, \ldots, S_N), \forall n \in \Nset
\end{align} of (\ref{eq:lamdba_nk}), where $\omega_{m}(S_m) = \frac{\sum_{k \in \Kset_{m}}  (\bar{p}_{m,k})^{\beta} \rho_{m,k}(S_{m}) }{\sum_{k \in \Kset_{m}} (\bar{p}_{m,k})^{\beta-1} }$. As seen, (\ref{eq:an}) is a convex function of $a_n$. 
Thus, we can define the region $\tilde{\Lambda} = \{ \{a_n\}_{n \in \Nset}|$ (\ref{eq:an}), (\ref{eq:sum_tau}),  $a_{n} \geq 0, \tau_{n}(S_1, \ldots, S_N) \geq 0, \forall n \in \Nset, (S_1, \ldots, S_{N}) \in \Sset \}$, which is clearly an inner bound on the actual stability region $\Lambda$. 
Despite the fact that $\tilde{\Lambda}$ is only inner bound on $\Lambda$, for some operating points, \ie at the intersection of  ${\lambda_{n,k}}/{\bar{p}_{n,k}} = {\lambda_{n,l}}/{\bar{p}_{n,l}}$, $k \neq l$ lines, the two stability regions ($\tilde{\Lambda}$ and $\Lambda$) coincide. Thus, for some utility functions, optimal operating points in both $\tilde{\Lambda}$ and $\Lambda$ coincide.
In the next section, we develop resource allocation schemes; $dFC$ and $qFC$ that achieve utility optimal operating points in $\tilde{\Lambda}$.

\section{Flow Control and Scheduling} \label{sec:oFC_qFC}
In this section, we develop resource allocation schemes; {\em deterministic FIFO-Control} ($dFC$), and a {\em queue-based FIFO control} ($qFC$). % to achieve utility optimal operating points in $\tilde{\Lambda}$. 

In general, our goal is to solve the optimization problem
\begin{align}\label{eq:main_opt}
\max_{\boldsymbol \lambda} \mbox{ } &  \sum_{n \in \Nset} \sum_{k \in \Kset_{n}} U_{n,k}(\lambda_{n,k}) \nonumber \\
\mbox{s.t.} \mbox{ }  & \lambda_{n,k} \in \Lambda, n \in \Nset, k \in \Kset_{n}
\end{align} and to find the corresponding optimal rates,  where $U_{n,k}$ is a concave utility function assigned to flow with rate $\lambda_{n,k}$. 
%The objective of the optimization problem in (\ref{eq:main_opt}) is to maximize the total utility $\sum_{n \in \Nset} \sum_{k \in \Kset_{n}} U_{n,k}(\lambda_{n,k})$. 
Although the objective function $\sum_{n \in \Nset} \sum_{k \in \Kset_{n}} U_{n,k}(\lambda_{n,k})$ in (\ref{eq:main_opt}) is concave, the optimization domain $\Lambda$ (\ie the stability region) may not be convex. Thus, we convert this problem to a convex optimization problem based on the structure of the inner bound we have developed in Section~\ref{sec:stability_single_innerBound}. 
In particular, setting $a_n = \lambda_{n,k}/(\bar{p}_{n,k})^{\beta}$, the problem in (\ref{eq:main_opt}) reduces to  $\max_{\boldsymbol a}$   $\sum_{n \in \Nset} \sum_{k \in \Kset_{n}} U_n(a_n(\bar{p}_{n,k})^{\beta})$, $a_{n} \in \tilde{\Lambda}, n \in \Nset$. This is our deterministic FIFO-control scheme; $dFC$ and expressed explicitly as;

\underline{Deterministic FIFO-Control ($dFC$):}
\begin{align} \label{eq:dFC}
\max_{\boldsymbol a, \tau} \mbox{ } &  \sum_{n \in \Nset} \sum_{k \in \Kset_{n}} U_{n,k}(a_n(\bar{p}_{n,k})^{\beta}) \nonumber \\
\mbox{s.t.} \mbox{ }  &  a_n \leq \sum_{(S_1, \ldots, S_N) \in \Sset} \frac{1_{[S_n]}}{\sum_{k \in \Kset_{n}} (\bar{p}_{n,k})^{\beta-1}} \nonumber \\
& \prod_{m \in \Nset - \{n\}} \omega_{m}(S_m)  \tau_{n}(S_1, \ldots, S_N), \forall n \in \Nset \nonumber \\
& \sum_{n \in \Nset} \tau_{n} (S_1, \ldots, S_N) \leq 1, \forall (S_1, \ldots, S_N) \in \Sset \nonumber \\
& a_n \geq 0, \forall n \in \Nset, (S_1, \ldots, S_N) \in \Sset \nonumber \\
& \tau_{n} (S_1, \ldots, S_N) \geq 0, \forall n \in \Nset, (S_1, \ldots, S_N) \in \Sset
\end{align} Note that $dFC$ optimizes $a_n$ and $\tau_{n}{(S_1, \ldots, S_N)}$. After the optimal values are determined, 
%$\lambda_{n,k}$ as well as packet transmission probability $\tau_{n}{(S_1, \ldots, S_N)}$ for each FIFO queue at every state. After the optimal values of $\lambda_{n,k}$ and $\tau_{(S_1, \ldots, S_N)}$ are determined, the 
packets are inserted into the FIFO queue $\Qset_{n}$ depending on $\lambda_{n,k} = a_n(\bar{p}_{n,k})^{\beta}$ and served from the FIFO queue $\Qset_{n}$ depending on $\tau_{n}{(S_1, \ldots, S_N)}$. 

Although $dFC$ gives us optimal operating points in the stability region; $\tilde{\Lambda}$, it is a centralized solution, and its adaptation to varying wireless channel conditions is limited. Thus, we also develop a more practical and queue-based FIFO-control scheme $qFC$, next. 

\underline{Queue-Based FIFO-Control ($qFC$):}
\begin{itemize} 
%\vspace{-5pt}
\item {\em Flow Control:} At every slot $t$, the flow controller attached to the FIFO queue $\Qset_{n}$ determines $a_n(t)$ according to; 
\begin{align} \label{eq:flow_control}
\max_{\boldsymbol a} \mbox{ } &  M  \bigl[\sum_{k \in \Kset_{n}} U_{n,k}(a_{n}(t) (\bar{p}_{n,k})^{\beta}) \bigr] - Q_{n}(t)a_{n}(t) \nonumber \\
\mbox{s.t.} \mbox{ }  &  a_n(t) \leq R_{n}^{max}, a_n(t) \geq 0 
\end{align} where $M$ is a large positive number, and $R_{n}^{max}$ is a positive value larger than the maximum outgoing rate from FIFO queue $\Qset_n$ (which is $R_{n}^{max} > 1$ as we assume that the maximum outgoing rate from a queue is 1 packet per slot). After $a_n(t)$ is determined according to (\ref{eq:flow_control}), $\lambda_{n,k}(t)$ is set as $\lambda_{n,k}(t)$ $=$ $a_{n}(t)$ $(\bar{p}_{n,k})^{\beta}$. Then, $\lambda_{n,k}(t)$ packets from the $k$th flow are inserted in $\Qset_{n}$. 
%\vspace{-5pt}
\item {\em Scheduling:} At slot $t$, the scheduling algorithm determines the FIFO queue from which a packet is transmitted according to;
\begin{align} \label{eq:scheduling}
\max_{\boldsymbol \tau} \mbox{ } &  \sum_{n \in \Nset} Q_{n}(t) \frac{1_{[S_{n}(t)]}}{\sum_{k \in \Kset_{n} } (\bar{p}_{n,k})^{\beta} }   \tau_{n}(S_1(t), \ldots, S_N(t)) \nonumber \\
\mbox{s.t.} \mbox{ }  &  \sum_{n \in \Nset} \tau_{n} (S_1(t), \ldots, S_N(t)) \leq 1, \nonumber \\
& \tau_{n} (S_1(t), \ldots, S_N(t)) \geq 0
\end{align} After $\tau_{n} (S_1(t), \ldots, S_N(t)) $ is determined, the outgoing traffic rate from queue $\Qset_{n}$ is set to $g_{n}(t)$ $=$ $\tau_{n}$ $(S_1(t),$ $\ldots,$ $S_N(t))$ $1_{[S_{n}(t)]}$, and $g_{n}(t)$ packets (which is 1 or 0 in our case) are transmitted from $\Qset_{n}$. 
\end{itemize}

Thus, the queue dynamics change according to (\ref{eq:queue_Qn}) and based on (\ref{eq:flow_control}) and (\ref{eq:scheduling}). Such queue dynamics lead to the following result. 
\begin{theorem} \label{theorem_lyap}
If the channel states are i.i.d. over time slots, the traffic arrival rates are controlled by the rate control algorithm in (\ref{eq:flow_control}), and the FIFO queues are served by the scheduling algorithm in (\ref{eq:scheduling}), then the admitted flow rates converge to the utility optimal operating point in the stability region $\tilde{\Lambda}$ with increasing $M$.
\end{theorem}
{\em Proof:} The proof is provided in Appendix B.
%Due to space limitations, the proof is provided in \cite{thisTechRep}.
\hfill $\Box$

\section{Performance Evaluation}\label{sec:performance}
In this section, we evaluate our $dFC$ and $qFC$ algorithms as compared to the baselines; (i) {\em optimal} solution, and (ii) {\em max-weight} algorithm for different number of FIFO queues and flows. Next, we briefly explain our baselines. 

\subsection{Baselines}
The {\em optimal} solution is a solution to (\ref{eq:main_opt}), and we compared $dFC$ and $qFC$ with the {\em optimal} solution for some scenarios where the stability region $\Lambda$ is convex. On the other hand, {\em max-weight} algorithm is a queue-based flow control and max-weight scheduling scheme. Our baseline {\em max-weight} algorithm mimics the structure of the solution provided in \cite{neely_mod}, and it is summarized briefly in the following. 

\underline{{\em Max-weight} for FIFO:}
\begin{itemize} 
%\vspace{-5pt}
\item {\em Flow Control:} At every time slot $t$, the flow controller attached to the FIFO queue $\Qset_{n}$ determines $\lambda_{n,k}(t)$ according to;
\begin{align} \label{eq:flow_control_mw}
\max_{\boldsymbol \lambda} \mbox{ } &  M  \bigl[\sum_{k \in \Kset_{n}} U_{n,k}(\lambda_{n,k}(t)) \bigr] - Q_{n,k}(t)\lambda_{n,k}(t) \nonumber \\
\mbox{s.t.} \mbox{ }  &  \lambda_{n,k}(t) \leq R_{n,k}^{max}, \forall k \in \Kset_{n}
\end{align} where $M$ and $R_{n,k}^{max}$ are positive large constants similar to (\ref{eq:flow_control}), and $Q_{n,k}(t)$ is the number of packets that belong to the $k$th flow in queue $\Qset_{n}$. 
\item {\em Scheduling:} At slot $t$, the scheduling algorithm determines the FIFO queue from which a packet is transmitted according to; 
\begin{align} \label{eq:scheduling_mw}
\max_{\boldsymbol \tau} \mbox{ } &  \sum_{n \in \Nset} Q_{n}(t) 1_{[S_{n}(t)]} \tau_{n}(S_1(t), \ldots, S_N(t)) \nonumber \\
\mbox{s.t.} \mbox{ }  &  \sum_{n \in \Nset} \tau_{n} (S_1(t), \ldots, S_N(t)) \leq 1 \nonumber \\
& \tau_{n} (S_1(t), \ldots, S_N(t)) \geq 0
\end{align} After $\tau_{n}$ $(S_1(t),$ $\ldots,$ $S_N(t)) $ is determined, a packet from the queue $\Qset_{n}$ is transmitted if $\tau_{n}$ $(S_1(t),$ $\ldots,$ $S_N(t))$ $=$ $1$; no packet is transmitted, otherwise. 
\end{itemize}
Next, we present our simulation results for single and multiple FIFO queues.

\subsection{Single-FIFO Queue}
In this section, we consider  a single FIFO queue $\Qset_{1}$. Similar to Section~{\ref{sec:stability_single_queue}}, we drop the queue index $n=1$ from the notation for brevity. In other words, we write $\lambda_{k}$ instead of $\lambda_{1,k}$, $p_{k}$ instead of $p_{1,k}$, and so on. 

\begin{figure}
%\vspace{-10pt}
\centering
\subfigure[Per-flow rates]{ \scalebox{.45}{\includegraphics{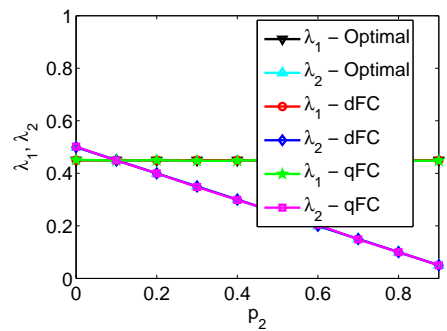}} } \hspace{-20pt}
\subfigure[Total rate]{ \scalebox{.45}{\includegraphics{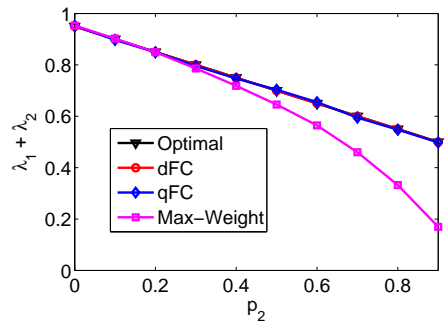}} }
\vspace{-10pt}
\caption{Single-FIFO queue shared by two flows when $p_1 = 0.1$, $\beta=1$, and $U_k(\lambda_k) = \log(\lambda_k)$. (a) Per-flow rates vs. $p_2$. (b) Total flow rate vs. $p_2$. }
\label{fig:sim_1}
\vspace{-5pt}
\end{figure}

Fig.~\ref{fig:sim_1} presents simulation results for a single queue and two flows for $p_1=0.1$, $\beta=1$, and $U_k(\lambda_k) = \log(\lambda_k)$. Fig.~\ref{fig:sim_1}(a) shows per-flow rates; $\lambda_1$ and $\lambda_2$ when $p_2$ is increasing. As seen, $\lambda_1$ is the same for all algorithms; optimal, $dFC$, and $qFC$. This also holds for $\lambda_2$. These results show that our algorithms $dFC$ and $qFC$ are as good as the optimal solution, and achieve the optimal operating points in $\Lambda$ in this scenario. The simulations results also show that our algorithms reduce the second flow rate $\lambda_2$ when $p_2$ increases while $\lambda_1$ and $p_1$ do not change. This means that our algorithms do not penalize a flow (flow 1) when the channel of another competing flow (flow 2) deteriorates, which shows the effectiveness of our algorithms to provide fairness. %to reduce the HOL blocking. 

Fig.~\ref{fig:sim_1}(b) shows the total rate $\lambda_1+\lambda_2$ versus $p_2$ for the same setup. As seen, our algorithms improves throughput over max-weight significantly. This is expected as our algorithms are designed to reduce the HOL blocking and to allocate wireless resources fairly among multiple flows. 

\begin{figure}
%\vspace{-10pt}
\centering
\subfigure[Flow rates]{ \scalebox{.45}{\includegraphics{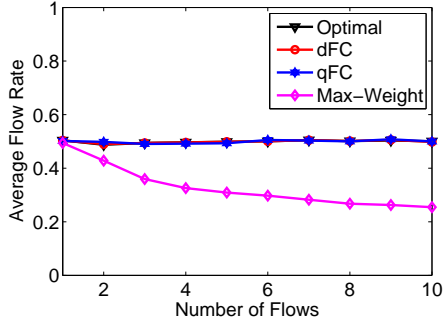}} } \hspace{-20pt}
\subfigure[Throughput improvement]{ \scalebox{.45}{\includegraphics{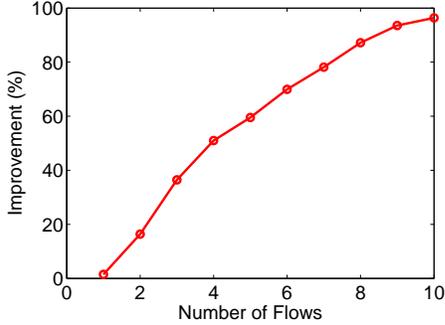}} }
\vspace{-10pt}
\caption{Single-FIFO queue shared by multiple flows. $p_k$ is selected randomly between $[0,1]$, $\beta = 1$, and $U_k(\lambda_k) = \log(\lambda_k)$. (a) Average flow rate versus number of flows. (b) Percentage of throughput improvement of $qFC$ over max-weight. }
\label{fig:sim_2}
\vspace{-5pt}
\end{figure}

Fig.~\ref{fig:sim_2} shows simulation results for a single queue shared by multiple flows. In this setup, $p_k$ is selected randomly between $[0,1]$, $\beta = 1$, $U_k(\lambda_k) = \log(\lambda_k)$. The simulations are repeated for 1000 different seeds, and the average values are reported.  Fig.~\ref{fig:sim_2}(a) shows average flow rate versus number of flows for our algorithms as well as max-weight. As seen, $dFC$ and $qFC$ are as good as the optimal solution, and they improve over max-weight significantly. Fig.~\ref{fig:sim_2}(b) shows the same simulation results, but reports the improvement of $qFC$ over max-weight. This figure shows that the improvement of our algorithms increases with increasing number of flows. Indeed, the improvement is up to 100\% when $K=10$, which is significant. The improvement is higher for large number of flows, because our algorithm allocates resources to the flows based on the quality of their channels and reduces the flow rate for the flows with bad channel conditions. However, max-weight does not have such a mechanism, and when there are more flows in the system, the probability of having a flow with bad channel condition increases, which reduces the overall throughput.

\subsection{Two-FIFO Queues}
In this section, we consider two FIFO queues $\Qset_{m}$ and $\Qset_{n}$. There are four flows in the system and each queue carries two flows, \ie $\Qset_{n}$ carries flows with rates $\lambda_{n,1}$, $\lambda_{n,2}$ and $\Qset_{m}$ carries flows with rates $\lambda_{m,1}$, $\lambda_{m,2}$. 

\begin{figure}
%\vspace{-10pt}
\centering
\subfigure{\scalebox{.45}{\includegraphics{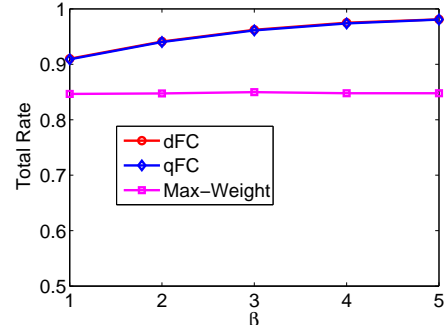}}}
\subfigure{\scalebox{.45}{\includegraphics{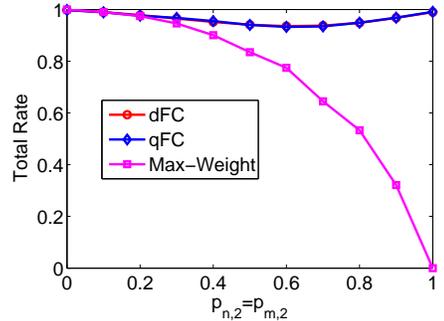}}}
\vspace{-10pt}
\caption{Two FIFO queues with four flows. (a) Total flow rate versus $\beta$ when $p_{n,1}=0.1$, $p_{n,1} = 0.5$, $p_{m,1} = 0.1$, $p_{m,2} = 0.5$, and $\log$ utility is employed, \ie $U_{n,k}(\lambda_{n,k}) = \log(\lambda_{n,k})$. (b) Total rate versus $p_{n,2}=p_{m,2}$ when $p_{n,1}=p_{m,1}=0.1$ and $\beta=2$. }
\label{fig:sim_3}
\vspace{-5pt}
\end{figure}

Fig.~(\ref{fig:sim_3})(a) shows the total flow rate versus $\beta$ for the scenario of two-FIFO queues with four flows when $p_{n,1}=0.1$, $p_{n,2} = 0.5$, $p_{m,1} = 0.1$, $p_{m,2} = 0.5$, and $\log$ utility is employed, \ie $U_{n,k}(\lambda_{n,k}) = \log(\lambda_{n,k})$. (We do not present the results of the optimal solution as the stability region $\Lambda$ is not convex in this scenario.) As seen, $dFC$ and $qFC$ have the same performance and improve over max-weight. The improvement increases with increasing $\beta$ as $dFC$ and $qFC$ penalize flows with bad channel conditions more when $\beta$ increases, which increases the total throughput.   

Fig.~(\ref{fig:sim_3})(b) shows the total rate versus $p_{n,2}=p_{m,2}$ for two-FIFO queues with four flows when $p_{n,1}=p_{m,1}=0.1$ and $\beta=2$. As seen, $dFC$ and $qFC$ improve significantly over max-weight. Furthermore, they achieve almost maximum achievable rate $1$ all the time. The reason is that $dFC$ and $qFC$ penalizes the queues with with bad channels. For example, when $p_{n,2}=p_{m,2}=1$, the total rate is $1$, because they allocate all the resources to $\lambda_{n,1}$ and $\lambda_{m,1}$ as there is no point to allocate those resources to $\lambda_{n,2}$ and $\lambda_{m,2}$ since their channels are always $OFF$. On the other hand, max-weight does not arrange the flow and queue service rates based on the channel conditions, so the total rate reduces to $0$ when $p_{n,2}=p_{m,2}=1$, \ie it is not possible to transmit any packets when max-weight is employed in this scenario.

%\begin{figure}
%%\vspace{-5pt}
%\centering
%\scalebox{.33}{\includegraphics{figs_matlab/two_queues_four_flows_rate_vs_beta.eps}}
%\caption{Total rate versus $\beta$ for the scenario of two-FIFO queues with four flows when $p_{n,1}=0.1$, $p_{n,1} = 0.5$, $p_{m,1} = 0.1$, $p_{m,2} = 0.5$, and $\log$ utility is employed, \ie $U_{n,k}(\lambda_{n,k}) = \log(\lambda_{n,k})$. }
%\label{fig:sim_3_a}
%\vspace{-10pt}
%\end{figure}
%
%\begin{figure}
%%\vspace{-5pt}
%\centering
%\scalebox{.33}{\includegraphics{figs_matlab/two_queues_four_flows_rate_vs_pn2_pm2.eps}}
%\caption{Total rate versus $p_{n,2}=p_{m,2}$ for the scenario of two-FIFO queues with four flows when $p_{n,1}=p_{m,1}=0.1$ and $\beta=2$. }
%\label{fig:sim_3_b}
%\vspace{-10pt}
%\end{figure}

\begin{figure}
%\vspace{-5pt}
\centering
\subfigure[$dFC$ and $qFC$]{ \scalebox{.45}{\includegraphics{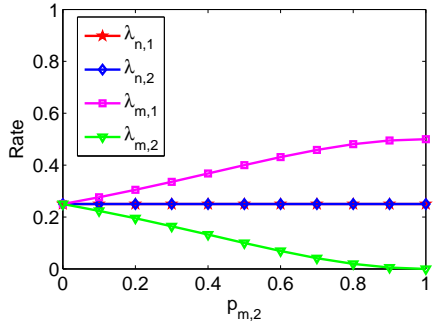}} } \hspace{-20pt}
\subfigure[Max-weight]{ \scalebox{.45}{\includegraphics{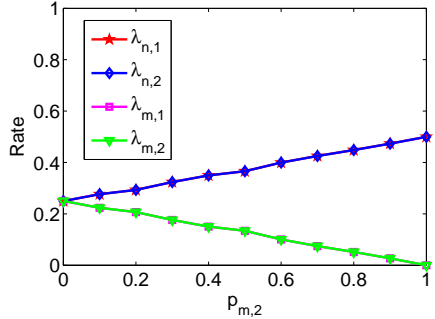}} }
\vspace{-10pt}
\caption{Per-flow rates versus $p_{m,2}$ for the scenario of two-FIFO queues with four flows when $p_{n,1}=p_{n,2}=p_{m,1}=0$ and $\beta=2$. (a) $dFC$ and $qFC$. (b) Max-weight.}
\label{fig:sim_4}
\vspace{-5pt}
\end{figure}

Fig.~\ref{fig:sim_4} further demonstrates how our algorithms treat flows with bad channel conditions. In particular, Fig.~\ref{fig:sim_4} presents per-flow rate versus $p_{m,2}$ for the scenario of two-FIFO queues with four flows when $p_{n,1}=p_{n,2}=p_{m,1}=0$ and $\beta=2$ for (a) $dFC$ and $qFC$ and (b) max-weight. As seen, when $p_{m,2}$ increases, $\lambda_{m,2}$ decreases in Fig.~\ref{fig:sim_4}(a) since its channel is getting worse. Yet, this does not affect the other flows. In fact, $\lambda_{m,1}$ even increases as more resources are allocated to it when $p_{m,2}$ increases. On the other hand, both $\lambda_{m,1}$ and $\lambda_{m,2}$ decrease with increasing $p_{m,2}$ in max-weight (Fig.~\ref{fig:sim_4}(b)). This is not fair, because $\lambda_{m,1}$ decreases with increasing $p_{m,2}$ although its channel is always $ON$ as $p_{m,1}=0$. In the same scenario (Fig.~\ref{fig:sim_4}(b)), the rates of the $n$th queue ($\lambda_{n,1}$ and $\lambda_{n,2}$) increase with increasing $p_{m,2}$ as they use available resource opportunistically. This makes the total rate the same for $dFC$, $qFC$, and max-weight. Yet, as we discussed, max-weight is not fair to flow $\lambda_{m,1}$ in this scenario. 

%as we mentioned max-weight penalized the first flow of $\Qset_{m}$ unfairly. 

\section{Related Work}\label{sec:related}
In this work, our goal is to understand FIFO queues in wireless networks and develop efficient flow control and scheduling policies for such a setup. In the seminal paper \cite{Karol87}, the authors analyze FIFO queues in an input queued switch. They show that the use of FIFO queues in that context limits the throughput to approximately 58\% of the maximum achievable throughput. However, in the context of wireless networks, similar results are in general not known.

Backpressure routing and scheduling framework has emer-ged from the pioneering work \cite{tass_eph1,tass_eph2}, which has generated a lot of research interest \cite{neely_book}; especially for wireless ad-hoc networks \cite{tass3,kahale,andrews,neely_mod_pow,stolyar_greedy,liu_stolyar}. Furthermore, it has been shown that backpressure can be combined with flow control to provide utility-optimal operation guarantee \cite{neely_mod,stolyar_greedy}. Such previous work mainly considered per-flow queues. However, FIFO queueing structure, which is the focus of this paper, is not compatible with the per-flow queueing requirements of these routing and scheduling schemes.

The strengths of backpressure-based network control have recently received increasing interest in terms of practical implementation. Multi-path TCP scheme is implemented over wireless mesh networks in \cite{horizon} for routing and scheduling packets using a backpressure based heuristic. At the link layer, \cite{DiffQ,umut_stolyar,sridharan2} propose, analyze, and evaluate link layer backpressure-based implementations with queue prioritization and congestion window size adjustment. Backpressure is implemented over sensor networks \cite{routing_wtht_routes} and wireless multi-hop networks \cite{xpress}. In these schemes, either last-in, first-out queueing is employed \cite{routing_wtht_routes} or link layer FIFO queues are strictly controlled \cite{xpress} to reduce the number of packets in the FIFO queues, hence HOL blocking.

In backpressure, each node constructs per-flow queues. There is some work in the literature to stretch this necessity. For example, \cite{pkt_by_pkt_adap_rout}, \cite{locbui} propose using real per-link and virtual per-flow queues. Such a method reduces the number of queues required in each node, and reduces the delay, but it still needs to construct per-link queues. Similarly, \cite{diffmax} constructs per-link queues in the link layer, and schedule packets according to FIFO rule from these queues. Such a setup is different than ours as per-link queues do not introduce HOL blocking. %On the other hand, we consider a scenario where constructing per-flow or per-link queues may not be feasible, and FIFO queues should be shared by multiple flows.

The main differences in our work are: (i) we consider FIFO queues shared by multiple flows where HOL blocking occurs as each flow is transmitted over a possibly different wireless link, (ii) we characterize the stability region of a general scenario where an arbitrary number of FIFO queues, which are served by a wireless medium, are shared by an arbitrary number of flows, and (iii) we develop efficient resource allocation schemes to exploit achievable rate in such a setup.

\section{Conclusion}\label{sec:conclusion}
We investigated the performance of FIFO queues over wireless networks and characterized the stability region of this system for arbitrary number of FIFO queues and flows. We developed inner bound on the stability region, and developed resource allocation schemes; $dFC$ and $qFC$, which achieve optimal operating point in the convex inner bound. Simulation results show that our algorithms significantly improve throughput in a wireless network with FIFO queues as compared to the well-known queue-based flow control and max-weight scheduling schemes.

\bibliographystyle{IEEEtran}

\section*{Appendix A: Proof of Theorem~\ref{theorem2}}
In this section, we provide a proof of Theorem~\ref{theorem2} for arbitrary number of FIFO queues and flows. 
Let us first consider $\lambda_{n,k}$, which should satisfy the following inequality. 
\begin{align} \label{eq:appA_1}
& \lambda_{n,k} \leq \sum_{(S_1,\ldots,S_N) \in \Sset} P[S_1,\ldots,S_N, H_n=k] 1_{[S_n]} \nonumber \\
& \tau_{n}(S_1,\ldots,S_N),  \forall n \in \Nset, k \in \Kset_{n}.
\end{align} where $P[S_1,\ldots,S_N, H_n=k]$ is the probability that the states of the queues are $(S_1,\ldots,S_N)$ and $H_n=k$, which is required as we can transmit a packet from the $k$th flow only when the HOL packet belongs to the $k$th flow. 
In this equation, we can calculate $P[S_1,\ldots,S_N, H_n=k]$ as 
\begin{align}
& P[S_1,\ldots,S_N, H_n=k]  = \sum_{l_1 \in \Kset_{1}} \ldots \sum_{l_{n-1} \in \Kset_{n-1}} \sum_{l_{n+1} \in \Kset_{n+1}} \ldots \nonumber \\
& \sum_{l_N \in \Kset_{N}} P[S_{1}, \ldots, S_{N}, H_{n}=k, H_1=l_1, \ldots, H_{n-1}=l_{n-1}, \nonumber \\
& H_{n+1}=l_{n+1}, \ldots, H_{N}=l_{N} ],
\end{align} 
\begin{align}
& P[S_1,\ldots,S_N, H_n=k]  = \sum_{l_1 \in \Kset_{1}} \ldots \sum_{l_{n-1} \in \Kset_{n-1}} \sum_{l_{n+1} \in \Kset_{n+1}} \ldots \nonumber \\
& \sum_{l_N \in \Kset_{N}} \underbrace{P[S_1 | H_1=l_1]}_{\triangleq \xi_{1,l_1}(S_1)} \ldots \underbrace{P[S_{n-1} | H_{n-1}=l_{n-1}]}_{\triangleq \xi_{n-1,l_{n-1}}(S_{n-1})} \nonumber \\
& \underbrace{P[S_{n} | H_{n}=k]}_{\triangleq \xi_{n,k}(S_{n})}  \underbrace{P[S_{n+1} | H_{n+1} = l_{n+1}]}_{\triangleq \xi_{n+1,l_{n+1}}(S_{n+1})} \ldots \underbrace{P[S_{N} | H_{N} = l_{N}]}_{\triangleq \xi_{N,l_{N}}(S_{N})} 
\nonumber \\
& P[H_{n}=k,  H_{1}=l_{1}, \ldots, H_{n-1}=l_{n-1}, H_{n+1}=l_{n+1}, \ldots, \nonumber \\
& H_{N}=l_{N}]
\end{align} Thus, we have
\begin{align} \label{eq:states_v1}
& P[S_1,\ldots,S_N, H_n=k]  = \sum_{l_1 \in \Kset_{1}} \ldots \sum_{l_{n-1} \in \Kset_{n-1}} \sum_{l_{n+1} \in \Kset_{n+1}} \ldots \nonumber \\
& \sum_{l_N \in \Kset_{N}} {\xi_{1,l_1}(S_1)} \ldots {\xi_{n-1,l_{n-1}}(S_{n-1})} {\xi_{n,k}(S_{n})} \nonumber \\
& {\xi_{n+1,l_{n+1}}(S_{n+1})}  \ldots {\xi_{N,l_{N}}(S_{N})} 
 P[H_{n}=k,  H_{1}=l_{1}, \ldots, \nonumber \\
& H_{n-1}=l_{n-1},   H_{n+1}=l_{n+1}, \ldots,  H_{N}=l_{N}]. 
\end{align} Now, we should calculate $P[H_{n}=k,  H_{1}=l_{1}, \ldots, H_{n-1}=l_{n-1}, H_{n+1}=l_{n+1}, \ldots,  H_{N}=l_{N}]$. 

We claim that $P[H_{n}=k,$  $H_{1}=l_{1},$ $\ldots,$ $H_{n-1}=l_{n-1},$ $H_{n+1}=l_{n+1},$ $\ldots,$  $H_{N}=l_{N}]$ $=$ $P[H_1=l_1]$ $\ldots$ $P[H_{n-1}=l_{n-1}]$ $P[H_n=k]$ $P[H_{n+1}=l_{n+1}]$ $\ldots$ $P[H_{N}=l_{N}]$. To prove this claim, we should show, without loosing generality, that the following conditions hold. 
\begin{align} \label{eq:app_cond_prob}
\mbox{C1: }  & P[H_{n}=k | H_1=l_1, \ldots, H_{n-1}=l_{n-1}, H_{n+1}=l_{n+1}, \nonumber \\
& \ldots, H_{N}=l_{N}] = P[H_{n}=k]  \nonumber \\ 
\mbox{C2: } & P[H_{n}=k | H_1=l_1, \ldots, H_{n-1}=l_{n-1}, H_{n+1}=l_{n+1}, \nonumber \\
& \ldots, H_{N-1}=l_{N-1}] = P[H_{n}=k]  \nonumber  \\ 
& \mbox{                } \vdots \nonumber \\
\mbox{CN: }  & P[H_{n}=k | H_1=l_1] = P[H_{n}=k] 
\end{align} We can calculate the conditional probabilities in the left hand side of the conditions; C1, C2, $\ldots$, CN in (\ref{eq:app_cond_prob}) by using a Markov chain. For C1, we can write a state transition probability of going from state $H_{n} = k$ to $H_{n}=m$ as $P[H_{n}=k \rightarrow H_{n}=m | H_1=l_1, \ldots, H_{n-1}=l_{n-1},$ $H_{n+1}=l_{n+1},$ $\ldots,$ $H_{N}=l_{N}]$, which is equal to $\bar{p}_{n,k} \pi_{n} \alpha_{n,k}$. \Ie $P[H_{n}=k \rightarrow H_{n}=m | H_1=l_1, \ldots,$ $H_{n-1}=l_{n-1},$ $H_{n+1}=l_{n+1}, \ldots, H_{N}=l_{N}]$ $=$ $\bar{p}_{n,k} \pi_{n} \alpha_{n,k}$. Similarly, if we write the state transition probabilities for the other conditions C2, $\ldots$, CN, we have $P[H_{n}=k \rightarrow H_{n}=m | H_1=l_1, \ldots, H_{n-1}=l_{n-1}, H_{n+1}=l_{n+1}, \ldots, H_{N}=l_{N}]$ $=$ $P[H_{n}=k \rightarrow H_{n}=m | H_1=l_1, \ldots, H_{n-1}=l_{n-1}, H_{n+1}=l_{n+1}, \ldots, H_{N-1}=l_{N-1}]$ $=$ $\ldots$ $=$ $P[H_{n}=k \rightarrow H_{n}=m | H_1=l_1]$ $=$ $P[H_{n}=k \rightarrow H_{n}=m]$ $=$ $P[H_{n}=k \rightarrow H_{n}=m]$ $=$ $\bar{p}_{n,k}\pi_{n}\alpha_{n,m}$. Therefore, in all Markov chains we can create for C1, C2, $\ldots$, CN, we have the same transition probabilities, so we have $P[H_{n}=k | H_1=l_1, \ldots, H_{n-1}=l_{n-1}, H_{n+1}=l_{n+1}, \ldots, H_{N}=l_{N}]$ $=$ $P[H_{n}=k | H_1=l_1, \ldots, H_{n-1}=l_{n-1}, H_{n+1}=l_{n+1}, \ldots, H_{N-1}=l_{N-1}]$ $=$ $P[H_{n}=k | H_1=l_1]$ $=$ $P[H_{n}=k]$. This proves our claim that $P[H_{n}=k,  H_{1}=l_{1}, \ldots,$ $H_{n-1}=l_{n-1},$ $H_{n+1}=l_{n+1}, \ldots,  H_{N}=l_{N}]$ $=$ $P[H_1=l_1]$ $\ldots$ $P[H_{n-1}=l_{n-1}]$ $P[H_n=k]$ $P[H_{n+1}=l_{n+1}]$ $\ldots$ $P[H_{N}=l_{N}]$. 

\begin{figure}
\vspace{5pt}
\centering
\scalebox{.45}{\includegraphics{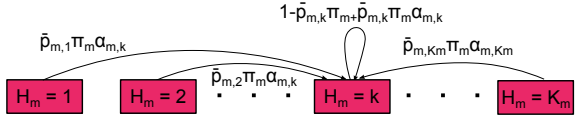}} 
\vspace{-5pt}
\caption{The state transition diagram for the states $H_{m}=k$, $\forall k \in \Kset_{m}$ and for the $m$th queue. Note that this state transition diagram only shows a subset of state transitions for clarity.}
\label{fig:app_Km_flows}
\vspace{-5pt}
\end{figure}

Now that we have shown that $P[H_{n}=k,  H_{1}=l_{1}, \ldots,$ $H_{n-1}=l_{n-1},$ $H_{n+1}=l_{n+1}, \ldots,  H_{N}=l_{N}]$ $=$ $P[H_1=l_1]$ $\ldots$ $P[H_{n-1}=l_{n-1}]$ $P[H_n=k]$ $P[H_{n+1}=l_{n+1}]$ $\ldots$ $P[H_{N}=l_{N}]$ holds, (\ref{eq:states_v1}) is expressed as
\begin{align} \label{eq:states_v2}
& P[S_1,\ldots,S_N, H_n=k]  = \sum_{l_1 \in \Kset_{1}} \ldots \sum_{l_{n-1} \in \Kset_{n-1}} \sum_{l_{n+1} \in \Kset_{n+1}} \ldots \nonumber \\
& \sum_{l_N \in \Kset_{N}} {\xi_{1,l_1}(S_1)}P[H_1=l_1] \ldots {\xi_{n-1,l_{n-1}}(S_{n-1})} \nonumber \\
&  P[H_{n-1}=l_{n-1}] {\xi_{n,k}(S_{n})} P[H_{n}=k] {\xi_{n+1,l_{n+1}}(S_{n+1})} \nonumber \\
&  P[H_{n+1}=l_{n+1}] \ldots {\xi_{N,l_{N}}(S_{N})} P[H_{N}=l_{N}],  
\end{align} 
%which is expressed as
%\begin{align} \label{eq:states_v3}
%& P[S_1,\ldots,S_N, H_n=k]  = \xi_{n,k}(S_n)P[H_n=k] \sum_{l_1 \in \Kset_{1}} \xi_{1,l_1}(S_1) \nonumber \\
%& P[H_1=l_1] \ldots \sum_{l_{n-1} \in \Kset_{n-1}} \xi_{n-1,l_{n-1}}(S_{n-1})P[H_{n-1}=l_{n-1}] \nonumber \\
%& \sum_{l_{n+1} \in \Kset_{n+1}} \xi_{n+1,l_{n+1}}(S_{n+1})P[H_{n+1}=l_{n+1}] \ldots \nonumber \\
%& \sum_{l_{N} \in \Kset_{N}} \xi_{N,l_{N}}(S_{N})  P[H_{N}=l_{N}], 
%\end{align} 
which leads to 
\begin{align} \label{eq:states_v4}
& P[S_1,\ldots,S_N, H_n=k]  =  \xi_{n,k}(S_n)P[H_n=k] \nonumber \\
& \prod_{m \in \Nset-\{n\}} \left( \sum_{k \in \Kset_{m}} \xi_{m,k}(S_{m}) P[H_{m}=k] \right). 
\end{align} Now, we should calculate $P[H_{m}=k]$ in (\ref{eq:states_v4}). The state transition diagram for the states $H_{m}=k$, $\forall k \in \Kset_{m}$ and for the $m$th queue is shown in Fig.~\ref{fig:app_Km_flows}. 
We can write the global balance equations for the state $H_m=k$ as 
\begin{align}
& P[H_m=1] \bar{p}_{m,1}\pi_{m}\alpha_{m,k} + \ldots + P[H_m=k][1 - \bar{p}_{m,k}\pi_{m} \nonumber \\
& + \bar{p}_{m,k}\pi_{m}\alpha_{m,k}] + \ldots + P[H_m=K_m] \bar{p}_{m,K_m}\pi_{m}\alpha_{m,k} = \nonumber \\
&  P[H_m=k], 
\end{align} which is expressed as
\begin{align}
& \pi_{m} \alpha_{m,k} ( \bar{p}_{m,1}P[H_m=1] + \ldots + \bar{p}_{m,k} P[H_m=k] + \ldots + \nonumber \\
& \bar{p}_{m,K_m} P[H_m=K_m]  ) = P[H_m = k] \bar{p}_{m,k} \pi_{m}, 
\end{align} which leads to 
\begin{align} \label{eq:app_state_mk}
& \alpha_{m,k} \sum_{i \in \Kset_{m}} \bar{p}_{m,i} P[H_m=i] = P[H_m=k] \bar{p}_{m,k}.  
\end{align} Similarly, the global balance equations for state $H_m=l$ leads to 
\begin{align} \label{eq:app_state_ml}
& \alpha_{m,l} \sum_{i \in \Kset_{m}} \bar{p}_{m,i} P[H_m=i] = P[H_m=l] \bar{p}_{m,l}.  
\end{align} From (\ref{eq:app_state_mk}) and (\ref{eq:app_state_ml}), we have 
\begin{align}
\frac{\alpha_{m,k}}{\alpha_{m,l}} = \frac{P[H_m=k]\bar{p}_{m,k}}{P[H_m=l]\bar{p}_{m,l}}.
\end{align} Thus, we have 
\begin{align}
P[H_m=l] = \frac{\alpha_{m,l}}{\bar{p}_{m,l}} \frac{P[H_m=k]}{\alpha_{m,k}} \bar{p}_{m,k}. 
\end{align} Since $\sum_{l \in \Kset_{m}} P[H_m = l] = 1$ should be satisfied, we have 
%\begin{align}
%\sum_{l \in \Kset_{m}-\{k\}} \left( \frac{\alpha_{m,l}}{\bar{p}_{m,l}} \frac{P[H_m=k]\bar{p}_{m,k}}{\alpha_{m,k}}  \right) + P[H_m = k] = 1
%\end{align}
%\begin{align}
%P[H_m = k] \left( \frac{\bar{p}_{m,k}}{\alpha_{m,k}} \left( \sum_{l \in \Kset_{m} - \{k\}} \frac{\alpha_{m,l}}{\bar{p}_{m,l}} \right) + 1 \right) = 1
%\end{align}
%\begin{align}
%P[H_m = k] \left( \left( \sum_{l \in \Kset_{m} - \{k\}} \frac{\alpha_{m,l}}{\bar{p}_{m,l}} \right) + \frac{\alpha_{m,k}}{\bar{p}_{m,k}} \right) = \frac{\alpha_{m,k}}{\bar{p}_{m,k}}
%\end{align}
%\begin{align}
%P[H_m = k] \left( \sum_{l \in \Kset_{m}} \frac{\alpha_{m,l}}{\bar{p}_{m,l}} \right)  = \frac{\alpha_{m,k}}{\bar{p}_{m,k}}
%\end{align}
%\begin{align}
%P[H_m = k] = \frac{\alpha_{m,k}/\bar{p}_{m,k}}{\sum_{l \in \Kset_{m}} \alpha_{m,l} / \bar{p}_{m,l}}
%\end{align}
\begin{align} \label{eq:app_prob_Hm_k}
P[H_m = k] = \frac{\lambda_{m,k}/\bar{p}_{m,k}}{\sum_{l \in \Kset_{m}} \lambda_{m,l} / \bar{p}_{m,l}}. 
\end{align} When (\ref{eq:app_prob_Hm_k}) is substituted in (\ref{eq:states_v4}), we have 
\begin{align} \label{eq:states_v5}
& P[S_1,\ldots,S_N, H_n=k]  = \xi_{n,k}(S_n) \frac{\lambda_{n,k}/\bar{p}_{n,k}}{\sum_{l \in \Kset_{m}} \lambda_{n,l} / \bar{p}_{n,l}} \nonumber \\
& \prod_{m \in \Nset-\{n\}} \frac{\sum_{k \in \Kset_{m}} \xi_{m,k}(S_m)\lambda_{m,k}/\bar{p}_{m,k}}{\sum_{k \in \Kset_{m}} \lambda_{m,k}/\bar{p}_{m,k}}. 
\end{align} Since $\rho_{m,k}(S_m) = \frac{\xi_{m,k}(S_m)}{\bar{p}_{m,k}}$, we have 
\begin{align} \label{eq:states_v6}
& P[S_1,\ldots,S_N, H_n=k]  = \xi_{n,k}(S_n) \frac{\lambda_{n,k}/\bar{p}_{n,k}}{\sum_{l \in \Kset_{m}} \lambda_{n,l} / \bar{p}_{n,l}} \nonumber \\
& \prod_{m \in \Nset-\{n\}} \frac{\sum_{k \in \Kset_{m}} \rho_{m,k}(S_m)\lambda_{m,k}}{\sum_{k \in \Kset_{m}} \lambda_{m,k}/\bar{p}_{m,k}}. 
\end{align} When we substitute (\ref{eq:states_v6}) into (\ref{eq:appA_1}), we have (\ref{eq:lamdba_nk}). This concludes the proof.

\section*{Appendix B: Proof of Theorem~\ref{theorem_lyap}}
Let define a Lyapunov function as; $L(\boldsymbol Q(t)) = \sum_{n \in \Nset} Q_{n}(t)$, and the Lyapunov drift as; $\Delta(\boldsymbol Q(t)) = E[L(\boldsymbol Q(t+1)) - L(\boldsymbol Q(t)) | \boldsymbol Q(t)]$, where $\boldsymbol Q(t) = \{Q_1(t), \ldots Q_{N}(t)\}$. Then, the Lyapunov drift is expressed as; 
\begin{align} \label{eq:appB_drift1}
\Delta(\boldsymbol Q(t)) = E[ \sum_{n \in \Nset} Q_{n}(t+1)^{2} - \sum_{n \in \Nset} Q_{n}(t)^2 | \boldsymbol Q(t)] 
\end{align}  

Note that we have, from Eq.~(\ref{eq:queue_Qn}) and the assumption $a_{n}(t) = \lambda_{n,k}(t)/(\bar{p}_{n,k})^{\beta}$ that,
\begin{align} \label{eq:appB_queue_Qn}
& Q_{n}(t+1) \leq \max [Q_{n}(t) - g_{n}(t), 0] + a_{n}(t) \sum_{k \in \Kset_{n}}(\bar{p}_{n,k})^{\beta}
\end{align} Using Eq.~(\ref{eq:appB_queue_Qn}) in Eq.~(\ref{eq:appB_drift1}), and using the fact that $(\max(Q-b,0)+A)^2 \leq Q^2 + A^2 + b^2 + 2Q(A-b)$, we have 
\begin{align}
& \Delta(\boldsymbol Q(t)) \leq E \bigl[ \sum_{n \in \Nset} \bigl\{ Q_{n}(t)^{2} + (a_{n}(t) \sum_{k \in \Kset_{n}} (\bar{p}_{n,k})^{\beta})^{2} + (g_{n}(t))^{2} \nonumber \\
& + 2Q_{n}(t) (a_n(t) \sum_{k \in \Kset_{n}} (\bar{p}_{n,k})^{\beta} - g_{n}(t) )  \bigr\}  - \sum_{n \in \Nset} Q_{n}(t)^{2} | \boldsymbol Q(t) \bigr]
\end{align} which is expressed as
\begin{align} \label{eq:appB_drift2}
& \frac{\Delta(\boldsymbol Q(t))}{2\sum_{k \in \Kset_{n}} (\bar{p}_{n,k})^{\beta} } \leq E \bigl[ \sum_{n \in \Nset} \bigl \{ \frac{(a_{n}(t))^{2}}{2} \sum_{k \in \Kset_{n}} (\bar{p}_{n,k})^{\beta} + \nonumber \\
& \frac{(g_{n}(t))^{2}}{2\sum_{k \in \Kset_{n}} (\bar{p}_{n,k})^{\beta}}  + Q_{n}(t) a_{n}(t) - \frac{Q_{n}(t) g_{n}(t)}{\sum_{k \in \Kset_{n}} (\bar{p}_{n,k})^{\beta} }  \bigr \} | \boldsymbol Q(t) \bigr]
\end{align} There always exist a finite and positive $B$ satisfying; $ B \geq E \bigl[ \sum_{n \in \Nset} \bigl \{ \frac{(a_{n}(t))^{2}}{2} \sum_{k \in \Kset_{n}} (\bar{p}_{n,k})^{\beta} + \frac{(g_{n}(t))^{2}}{2\sum_{k \in \Kset_{n}} (\bar{p}_{n,k})^{\beta}} \bigr \} \bigr ]$. Thus, Eq.~(\ref{eq:appB_drift2}) is expressed as; 
\begin{align} \label{eq:appB_drift3}
& \frac{\Delta(\boldsymbol Q(t))}{2\sum_{k \in \Kset_{n}} (\bar{p}_{n,k})^{\beta} } \leq B + E \bigl[ \sum_{n \in \Nset} Q_{n}(t) \bigl(a_{n}(t) - \nonumber \\
& \frac{g_{n}(t)}{\sum_{k \in \Kset_{n}} (\bar{p}_{n,k})^{\beta} } \bigr) | \boldsymbol Q(t)\bigr]
\end{align} Note that if the flow arrival rates $\lambda_{n,k}(t) = a_{n}(t) (\bar{p}_{n,k})^{\beta}$ are inside the capacity region $\tilde{\Lambda}$, then the minimizing the right hand side of the drift inequality in Eq.~(\ref{eq:appB_drift3}) corresponds to the scheduling part of $qFC$ in Eq.~(\ref{eq:scheduling}). 

Now, let us consider again the stability region constraint in Eq.~(\ref{eq:appA_1}), which is $\lambda_{n,k}$ $\leq$ $\sum_{(S_1,\ldots,S_N) \in \Sset} P[S_1,\ldots,S_N, H_n=k] 1_{[S_n]}  \tau_{n}(S_1,\ldots,S_N),  \forall n \in \Nset, k \in \Kset_{n}$, and expressed as; 
\begin{align}
& \sum_{k \in \Kset_{n}} \lambda_{n,k} \leq \sum_{(S_1 \ldots S_N) \in \Sset} \bigl( \sum_{k \in \Kset_{n}} P[S_1 \ldots S_N, H_n=k]  \bigr) 1_{[S_n]} \nonumber \\
& \tau_{n}(S_1 \ldots S_N)
\end{align} which is equal to
\begin{align}
\sum_{k \in \Kset_{n}} \lambda_{n,k} \leq \sum_{(S_1 \ldots S_N) \in \Sset}  P[S_1 \ldots S_N]   1_{[S_n]} \tau_{n}(S_1 \ldots S_N)
\end{align} Since $\lambda_{n,k} = a_n (\bar{p}_{n,k})^{\beta}$, we have
\begin{align}
\sum_{k \in \Kset_{n}} a_{n} (\bar{p}_{n,k})^{\beta} \leq \sum_{(S_1 \ldots S_N) \in \Sset}  P[S_1 \ldots S_N]   1_{[S_n]} \tau_{n}(S_1 \ldots S_N)
\end{align}
\begin{align}
a_{n} \sum_{k \in \Kset_{n}}  (\bar{p}_{n,k})^{\beta} \leq \sum_{(S_1 \ldots S_N) \in \Sset}  P[S_1 \ldots S_N]   1_{[S_n]} \tau_{n}(S_1 \ldots S_N)
\end{align}
 \begin{align} \label{eq:appB_gec1}
a_{n}  \leq \sum_{(S_1 \ldots S_N) \in \Sset}  P[S_1 \ldots S_N]   \frac{1_{[S_n]}\tau_{n}(S_1 \ldots S_N)}{\sum_{k \in \Kset_{n}}  (\bar{p}_{n,k})^{\beta}} 
\end{align} Let $g_n = 1_{[S_n]}\tau_{n}(S_1 \ldots S_N)$. Then, Eq.~(\ref{eq:appB_gec1}) is expressed as; 
 \begin{align}
a_{n}  \leq \sum_{(S_1 \ldots S_N) \in \Sset}  P[S_1 \ldots S_N]   \frac{g_n}{\sum_{k \in \Kset_{n}}  (\bar{p}_{n,k})^{\beta}} 
\end{align} There exists a small positive value $\epsilon$ satisfying
 \begin{align}
a_{n} + \epsilon \leq \sum_{(S_1 \ldots S_N) \in \Sset}  P[S_1 \ldots S_N]   \frac{g_n}{\sum_{k \in \Kset_{n}}  (\bar{p}_{n,k})^{\beta}} 
\end{align} Thus, we can find a randomized policy satisfying 
\begin{align} \label{eq:appB_randEp}
E \bigl[ \oset{*}{a}_n(t) - \frac{\oset{*}{g}_n(t)}{\sum_{k \in \Kset_{n}}  (\bar{p}_{n,k})^{\beta}} \bigr] \leq -\epsilon
\end{align}

Now, let us consider Eq.~(\ref{eq:appB_drift3}) again, which is expressed as; 
\begin{align}
& \frac{\Delta(\boldsymbol Q(t))}{2\sum_{k \in \Kset_{n}} (\bar{p}_{n,k})^{\beta} } \leq B + \sum_{n \in \Nset} Q_{n}(t) E \bigl[  a_{n}(t) - \nonumber \\
& \frac{g_{n}(t)}{\sum_{k \in \Kset_{n}} (\bar{p}_{n,k})^{\beta} }  | \boldsymbol Q(t)\bigr]
\end{align} We minimize the right hand side of Eq.~(\ref{eq:appB_drift3}), so the following inequality satisfies; 
\begin{align} \label{eq:appB_drift4}
& E \bigl[  a_{n}(t) -  \frac{g_{n}(t)}{\sum_{k \in \Kset_{n}} (\bar{p}_{n,k})^{\beta} }  | \boldsymbol Q(t)\bigr] \leq E \bigl[  \oset{*}{a}_{n}(t) -  \frac{\oset{*}{g}_{n}(t)}{\sum_{k \in \Kset_{n}} (\bar{p}_{n,k})^{\beta} } \nonumber \\
& | \boldsymbol Q(t)\bigr]
\end{align} where $\oset{*}{a}_{n}(t)$ and $\oset{*}{g}_{n}(t)$ are the solutions of a randomized policy. Incorporating Eq.~(\ref{eq:appB_randEp}) in Eq.~(\ref{eq:appB_drift4}), we have
\begin{align}\label{eq:appB_drift5}
\frac{\Delta(\boldsymbol Q(t))}{2\sum_{k \in \Kset_{n}} (\bar{p}_{n,k})^{\beta} } \leq B - \epsilon \sum_{n \in \Nset} Q_{n}(t)
\end{align} The time average of Eq.~(\ref{eq:appB_drift5}) leads to
\begin{align}
& \limsup_{t \rightarrow \infty}  \frac{1}{t} \sum_{\tau = 0}^{t-1} \frac{\Delta(\boldsymbol Q(\tau))}{2 \sum_{k \in \Kset_{n}} (\bar{p}_{n,k})^{\beta} } \leq \limsup_{t \rightarrow \infty}  \frac{1}{t} \sum_{\tau = 0}^{t-1} \bigl[ B - \nonumber \\
& \epsilon \sum_{n \in \Nset} Q_{n}(\tau) \bigr]
\end{align} 
\begin{align}
\limsup_{t \rightarrow \infty}  \frac{1}{t} \sum_{\tau = 0}^{t-1} \bigl( \sum_{n \in \Nset} Q_{n}(\tau) \bigr) \leq \frac{B}{\epsilon}
\end{align} This concludes that the time average of the queues are bounded if the arrival rates are inside the capacity region $\tilde{\Lambda}$. 

Now, let us focus on the original claim of Theorem~\ref{theorem2}. Let us consider a drift+penalty function as; 
\begin{align}
& \frac{\Delta(\boldsymbol Q(t))}{2\sum_{k \in \Kset_{n}} (\bar{p}_{n,k})^{\beta} } - \sum_{n \in \Nset} \sum_{k \in \Kset_{n}} M E[U_{n,k} (\lambda_{n,k}(t)) | \boldsymbol Q(t) ] 
\leq \nonumber \\
& B + E \bigl[ \sum_{n \in \Nset} Q_{n}(t) \bigl(a_{n}(t) -  \frac{g_{n}(t)}{\sum_{k \in \Kset_{n}} (\bar{p}_{n,k})^{\beta} } \bigr) | \boldsymbol Q(t)\bigr] - \nonumber \\
& \sum_{n \in \Nset} \sum_{k \in \Kset_{n}} M E[U_{n,k} (\lambda_{n,k}(t)) | \boldsymbol Q(t) ] 
\end{align} Since we set $\lambda_{n,k}(t) = a_{n}(t) (\bar{p}_{n,k})^{\beta}$, we have
\begin{align} \label{eq:appB_dpp1}
& \frac{\Delta(\boldsymbol Q(t))}{2\sum_{k \in \Kset_{n}} (\bar{p}_{n,k})^{\beta} } - \sum_{n \in \Nset} \sum_{k \in \Kset_{n}} M E[U_{n,k} (a_{n}(t) (\bar{p}_{n,k})^{\beta}) | \boldsymbol Q(t) ] 
\nonumber \\
& \leq  B + \sum_{n \in \Nset} E \bigl[  Q_{n}(t) \bigl(a_{n}(t) -  \frac{g_{n}(t)}{\sum_{k \in \Kset_{n}} (\bar{p}_{n,k})^{\beta} } \bigr) | \boldsymbol Q(t)\bigr] - \nonumber \\
& \sum_{n \in \Nset} \sum_{k \in \Kset_{n}} M E[U_{n,k} (a_{n}(t) (\bar{p}_{n,k})^{\beta}) | \boldsymbol Q(t) ] 
\end{align} Note that minimizing the right hand side of Eq.~(\ref{eq:appB_dpp1}) corresponds to the flow control and scheduling algorithms of $qFC$ in Eq.~(\ref{eq:flow_control}) and Eq.~(\ref{eq:scheduling}), respectively. Since there exists a randomized policy satisfying Eq.~(\ref{eq:appB_randEp}), Eq.~(\ref{eq:appB_dpp1}) is expressed as
\begin{align} \label{eq:appB_dpp2}
& \frac{\Delta(\boldsymbol Q(t))}{2\sum_{k \in \Kset_{n}} (\bar{p}_{n,k})^{\beta} } - \sum_{n \in \Nset} \sum_{k \in \Kset_{n}} M E[U_{n,k} (a_{n}(t) (\bar{p}_{n,k})^{\beta}) | \boldsymbol Q(t) ] 
\nonumber \\
& \leq  B - \epsilon \sum_{n \in \Nset} Q_{n}(t) - \sum_{n \in \Nset } \sum_{k \in \Kset_{n}} M U_{n,k}(A_n(\bar{p}_{n,k})^{\beta} + \delta) 
\end{align} where $\sum_{n \in \Nset} \sum_{k \in \Kset_{n}} U_{n,k}(A_n(\bar{p}_{n,k})^{\beta} + \delta)$ is the maximum time average of the sum utility function that can be achieved by any control policy that stabilizes the system. Then, the time average of Eq.~(\ref{eq:appB_dpp2}) becomes 
\begin{align} \label{eq:appB_dpp3}
& \limsup_{t \rightarrow \infty}  \frac{1}{t} \sum_{\tau = 0}^{t-1} \biggl \{ \frac{\Delta(Q(\tau))}{2 \sum_{k \in \Kset_{n}} (\bar{p}_{n,k})^{\beta} }  - \nonumber \\
& \sum_{n \in \Nset} \sum_{k \in \Kset_{n}} M E[U_{n,k} (a_{n}(\tau) (\bar{p}_{n,k})^{\beta}) | \boldsymbol Q(t) ]   \biggr \} \leq  \nonumber \\
& \limsup_{t \rightarrow \infty}  \frac{1}{t} \sum_{\tau = 0}^{t-1}  \biggl \{ B - \epsilon \sum_{n \in \Nset} Q_{n}(\tau) - \nonumber \\
& \sum_{n \in \Nset} \sum_{k \in \Kset_{n}} M U_{n,k} (A_n(\bar{p}_{n,k})^{\beta} + \delta)  \biggr \}
\end{align} Now, let us first consider the stability of the queues. If both sides of Eq.~(\ref{eq:appB_dpp3}) is divided by $\epsilon$ and the terms are arranged, we have
\begin{align}
& \limsup_{t \rightarrow \infty}  \frac{1}{t} \sum_{\tau = 0}^{t-1} \bigl \{ \sum_{n \in \Nset} Q_{n}(\tau) \bigr \} \leq \frac{B}{\epsilon} + \limsup_{t \rightarrow \infty}  \frac{1}{t} \sum_{\tau = 0}^{t-1} \bigl \{ \sum_{n \in \Nset} \sum_{k \in \Kset_{n}} \nonumber \\
& \frac{M}{\epsilon} E[U_{n,k}(a_{n}(\tau) (\bar{p}_{n,k})^{\beta} )] \bigr \} - \sum_{k \in \Nset} \sum_{k \in \Kset_{n}} \frac{M}{\epsilon} U_{n,k} (A_n(\bar{p}_{n,k})^{\beta} + \delta)
\end{align} Since the right hand side is a positive finite value, this concludes that the time averages of the total queue sizes are bounded. 

Now, let us consider the optimality. If both sides of Eq.~(\ref{eq:appB_dpp3}) are divided by $M$, we have
\begin{align}
& - \limsup_{t \rightarrow \infty}  \frac{1}{t} \sum_{\tau = 0}^{t-1}  \sum_{n \in \Nset} \sum_{k \in \Kset_{n}} E[U_{n,k} (a_{n}(\tau) (\bar{p}_{n,k})^{\beta} ) ] \leq  \nonumber \\
& \limsup_{t \rightarrow \infty}  \frac{1}{t} \sum_{\tau = 0}^{t-1}  \bigl \{ \frac{B}{M} - \frac{\epsilon}{M} \sum_{n \in \Nset} Q_{n}(\tau) - \nonumber \\
& \sum_{n \in \Nset} \sum_{k \in \Kset_{n}}  U_{n,k} (A_n(\bar{p}_{n,k})^{\beta} + \delta)  \bigr \}
\end{align} By arranging the terms, we have 
\begin{align}
& \limsup_{t \rightarrow \infty}  \frac{1}{t} \sum_{\tau = 0}^{t-1}  \sum_{n \in \Nset} \sum_{k \in \Kset_{n}} E[U_{n,k} (a_{n}(\tau) (\bar{p}_{n,k})^{\beta} ) ] \geq  \nonumber \\
& \limsup_{t \rightarrow \infty}  \frac{1}{t} \sum_{\tau = 0}^{t-1}  \bigl \{  \sum_{n \in \Nset} \sum_{k \in \Kset_{n}}  U_{n,k} (A_n(\bar{p}_{n,k})^{\beta} + \delta) - \frac{B}{M} \nonumber \\
& + \frac{\epsilon}{M} \sum_{n \in \Nset} Q_{n}(\tau)  \bigr \}
\end{align} Since $\frac{\epsilon}{M} \sum_{n \in \Nset} Q_{n}(\tau)$ is positive for any $\tau$, we have
\begin{align}
& \limsup_{t \rightarrow \infty}  \frac{1}{t} \sum_{\tau = 0}^{t-1}  \sum_{n \in \Nset} \sum_{k \in \Kset_{n}} E[U_{n,k} (a_{n}(\tau) (\bar{p}_{n,k})^{\beta} ) ] \geq  \nonumber \\
& \limsup_{t \rightarrow \infty}  \frac{1}{t} \sum_{\tau = 0}^{t-1}  \bigl \{  \sum_{n \in \Nset} \sum_{k \in \Kset_{n}}  U_{n,k} (A_n(\bar{p}_{n,k})^{\beta} + \delta) - \frac{B}{M}   \bigr \}
\end{align} which leads to
\begin{align}
& \limsup_{t \rightarrow \infty}  \frac{1}{t} \sum_{\tau = 0}^{t-1}  \sum_{n \in \Nset} \sum_{k \in \Kset_{n}} E[U_{n,k} (a_{n}(\tau) (\bar{p}_{n,k})^{\beta} ) ] \geq  \nonumber \\
& \sum_{n \in \Nset} \sum_{k \in \Kset_{n}}  U_{n,k} (A_n(\bar{p}_{n,k})^{\beta} + \delta) - \frac{B}{M} 
\end{align} This proves that the admitted flow rates converge to the utility optimal operating point with increasing $M$. This concludes the proof.

\end{document}